\documentclass{article}
\usepackage{graphicx}
\usepackage{amsfonts}

\begin{document}

\title{Newtonian gravity, red shift, confinement, asymptotic freedom and
quarks oscillations }
\author{G. Quznetsov \footnote{e-mails: gunn@mail.ru, quznets@yahoo.com,gunnqu@gmail.com}\\
Chelyabinsk State University}
\maketitle

\begin{abstract}
Quarks oscillations give the Newtonian gravity law, the red shift, the
confinement and the asymptotic freedom.
\end{abstract}

Let $\left\langle \rho \left( \underline{x}\right) ,j_1\left( \underline{x}%
\right) ,j_2\left( \underline{x}\right) ,j_3\left( \underline{x}\right)
\right\rangle =\left\langle \rho \left(t,\mathbf{x}\right) ,\mathbf{j}\left(
t,\mathbf{x}\right) \right\rangle $ be a probability density 3+1-vector of
any physical event \cite{Q0}.

Complex functions $\varphi _1\left( \underline{x}\right) $, $\varphi
_2\left( \underline{x}\right) $, $\varphi _3\left( \underline{x}\right) $, $%
\varphi _4\left( \underline{x}\right) $ exist \cite{Q200} such that

\begin{eqnarray}
\rho &=&\sum_{s=1}^4\varphi _s^{*}\varphi _s\mbox{,}  \label{j} \\
\frac{j_{\alpha}}{\mathrm{c}} &=&-\sum_{k=1}^4\sum_{s=1}^4\varphi
_s^{*}\beta _{s,k}^{\left[ \alpha \right] }\varphi _k  \nonumber
\end{eqnarray}

for every such density vector (here $\alpha \in \{1,2,3\}$,

$\beta ^{\left[ 1\right] }:=\left[ 
\begin{array}{cccc}
0 & 1 & 0 & 0 \\ 
1 & 0 & 0 & 0 \\ 
0 & 0 & 0 & -1 \\ 
0 & 0 & -1 & 0
\end{array}
\right] $, $\beta ^{\left[ 2\right] }:=\left[ 
\begin{array}{cccc}
0 & -\mathrm{i} & 0 & 0 \\ 
\mathrm{i} & 0 & 0 & 0 \\ 
0 & 0 & 0 & \mathrm{i} \\ 
0 & 0 & -\mathrm{i} & 0
\end{array}
\right] $,

$\beta ^{\left[ 3\right] }:=\left[ 
\begin{array}{cccc}
1 & 0 & 0 & 0 \\ 
0 & -1 & 0 & 0 \\ 
0 & 0 & -1 & 0 \\ 
0 & 0 & 0 & 1
\end{array}
\right] $).

If

\[
\varphi =\left[ 
\begin{array}{c}
\varphi _1 \\ 
\varphi _2 \\ 
\varphi _3 \\ 
\varphi _4
\end{array}
\right] 
\]

then \cite{Q1}:

\begin{equation}
\frac 1{\mathrm{c}}\partial _t\varphi +\left( \mathrm{i}\Theta _0+\mathrm{i}%
\Upsilon _0\gamma ^{\left[ 5\right] }\right) \varphi =\left( 
\begin{array}{c}
\beta ^{\left[ 1\right] }\partial _1+\mathrm{i}\Theta _1\beta ^{\left[
1\right] }+\mathrm{i}\Upsilon _1\beta ^{\left[ 1\right] }\gamma ^{\left[
5\right] }+ \\ 
+\beta ^{\left[ 2\right] }\partial _2+\mathrm{i}\Theta _2\beta ^{\left[
2\right] }+\mathrm{i}\Upsilon _2\beta ^{\left[ 2\right] }\gamma ^{\left[
5\right] }+ \\ 
+\beta ^{\left[ 3\right] }\partial _3+\mathrm{i}\Theta _3\beta ^{\left[
3\right] }+\mathrm{i}\Upsilon _3\beta ^{\left[ 3\right] }\gamma ^{\left[
5\right] }+ \\ 
+\mathrm{i}M_0\gamma ^{\left[ 0\right] }+\mathrm{i}M_4\beta ^{\left[
4\right] }- \\ 
-\mathrm{i}M_{\zeta ,0}\gamma _\zeta ^{[0]}+\mathrm{i}M_{\zeta ,4}\zeta
^{[4]}- \\ 
-\mathrm{i}M_{\eta ,0}\gamma _\eta ^{[0]}-\mathrm{i}M_{\eta ,4}\eta ^{[4]}+
\\ 
+\mathrm{i}M_{\theta ,0}\gamma _\theta ^{[0]}+\mathrm{i}M_{\theta ,4}\theta
^{[4]}
\end{array}
\right) \varphi \mbox{.}  \label{ham0}
\end{equation}
with real $\Theta _k$, $\Upsilon _k$, $M_0$, $M_4$, $M_{\zeta ,0}$, $%
M_{\zeta ,4}$, $M_{\eta ,0}$, $M_{\eta ,4}$, $M_{\theta ,0}$, $M_{\theta ,4}$
and 
\begin{equation}
\gamma ^{\left[ 5\right] }\stackrel{def}{=}\left[ 
\begin{array}{cc}
1_2 & 0_2 \\ 
0_2 & -1_2
\end{array}
\right] \mbox{,}  \label{g5}
\end{equation}

From \cite{Q000}: {\it the colored motion equation} as the following:

\begin{equation}
\left( 
\begin{array}{c}
-\beta ^{\left[ 0\right] }\mathrm{i}\partial _0+\Theta _0\beta ^{\left[
0\right] }+\Upsilon _0\beta ^{\left[ 0\right] }\gamma ^{\left[ 5\right] }+
\\ 
-\beta ^{\left[ 1\right] }\mathrm{i}\partial _1+\Theta _1\beta ^{\left[
1\right] }+\Upsilon _1\beta ^{\left[ 1\right] }\gamma ^{\left[ 5\right] }-
\\ 
-\beta ^{\left[ 2\right] }\mathrm{i}\partial _2+\Theta _2\beta ^{\left[
2\right] }+\Upsilon _2\beta ^{\left[ 2\right] }\gamma ^{\left[ 5\right] }-
\\ 
-\beta ^{\left[ 3\right] }\mathrm{i}\partial _3+\Theta _3\beta ^{\left[
3\right] }+\Upsilon _3\beta ^{\left[ 3\right] }\gamma ^{\left[ 5\right] }-
\\ 
-M_{\zeta ,0}\gamma _\zeta ^{[0]}+M_{\zeta ,4}\zeta ^{[4]}+ \\ 
-M_{\eta ,0}\gamma _\eta ^{[0]}-M_{\eta ,4}\eta ^{[4]}+ \\ 
+M_{\theta ,0}\gamma _\theta ^{[0]}+M_{\theta ,4}\theta ^{[4]}
\end{array}
\right) \varphi =0\mbox{.}  \label{clrH}
\end{equation}

Here \cite{PT}:

\[
\gamma _\zeta ^{[0]}=\left[ 
\begin{array}{cccc}
0 & 0 & 0 & -1 \\ 
0 & 0 & -1 & 0 \\ 
0 & -1 & 0 & 0 \\ 
-1 & 0 & 0 & 0
\end{array}
\right] ,\zeta ^{[4]}=\left[ 
\begin{array}{cccc}
0 & 0 & 0 & \mathrm{i} \\ 
0 & 0 & \mathrm{i} & 0 \\ 
0 & -\mathrm{i} & 0 & 0 \\ 
-\mathrm{i} & 0 & 0 & 0
\end{array}
\right] 
\]

are the antidiagonal (mass) elements of red pentad;

\[
\gamma _\eta ^{[0]}=\left[ 
\begin{array}{cccc}
0 & 0 & 0 & \mathrm{i} \\ 
0 & 0 & -\mathrm{i} & 0 \\ 
0 & \mathrm{i} & 0 & 0 \\ 
-\mathrm{i} & 0 & 0 & 0
\end{array}
\right] ,\eta ^{[4]}=\left[ 
\begin{array}{cccc}
0 & 0 & 0 & 1 \\ 
0 & 0 & -1 & 0 \\ 
0 & -1 & 0 & 0 \\ 
1 & 0 & 0 & 0
\end{array}
\right] 
\]

are the antidiagonal (mass) elements of green pentad;

\[
\gamma _\theta ^{[0]}=\left[ 
\begin{array}{cccc}
0 & 0 & -1 & 0 \\ 
0 & 0 & 0 & 1 \\ 
-1 & 0 & 0 & 0 \\ 
0 & 1 & 0 & 0
\end{array}
\right] ;\theta ^{[4]}=\left[ 
\begin{array}{cccc}
0 & 0 & -\mathrm{i} & 0 \\ 
0 & 0 & 0 & \mathrm{i} \\ 
-\mathrm{i} & 0 & 0 & 0 \\ 
0 & \mathrm{i} & 0 & 0
\end{array}
\right] 
\]

are the antidiagonal (mass) elements of blue pentad;

$M_{\zeta ,0}$, $M_{\zeta ,4}$ are red down and up quark's mass numbers;

$M_{\eta ,0}$, $M_{\eta ,4}$ are green down and up quark's mass numbers;

$M_{\theta ,0}$, $M_{\theta ,4}$ are blue down and up quark's mass numbers.

The quarks mass members of this equation form the following matrix sum:

\[
\widehat{M}\stackrel{def}{=} 
\begin{array}{c}
-M_{\zeta ,0}\gamma _\zeta ^{[0]}+M_{\zeta ,4}\zeta ^{[4]}- \\ 
-M_{\eta ,0}\gamma _\eta ^{[0]}-M_{\eta ,4}\eta ^{[4]}+ \\ 
+M_{\theta ,0}\gamma _\theta ^{[0]}+M_{\theta ,4}\theta ^{[4]}
\end{array}
\]

\[
=\left[ 
\begin{array}{cccc}
0 & 0 & -M_{\theta ,0} & M_{\zeta ,0}-\mathrm{i}M_{\eta ,0} \\ 
0 & 0 & M_{\zeta ,0}+\mathrm{i}M_{\eta ,0} & M_{\theta ,0} \\ 
-M_{\theta ,0} & M_{\zeta ,0}-\mathrm{i}M_{\eta ,0} & 0 & 0 \\ 
M_{\zeta ,0}+\mathrm{i}M_{\eta ,0} & M_{\theta ,0} & 0 & 0
\end{array}
\right] 
\]

\[
+\mathrm{i}\left[ 
\begin{array}{cccc}
0 & 0 & -M_{\theta ,4} & M_{\zeta ,4}+\mathrm{i}M_{\eta ,4} \\ 
0 & 0 & M_{\zeta ,4}-\mathrm{i}M_{\eta ,4} & M_{\theta ,4} \\ 
-M_{\theta ,4} & -M_{\zeta ,4}-\mathrm{i}M_{\eta ,4} & 0 & 0 \\ 
-M_{\zeta ,4}+\mathrm{i}M_{\eta ,4} & M_{\theta ,4} & 0 & 0
\end{array}
\right] \mbox{.} 
\]

Elements of these matrices can be turned by formula of shape \cite{Z}:

\[
=\left( 
\begin{array}{cc}
\cos \frac \theta 2 & \mathrm{i}\sin \frac \theta 2 \\ 
\mathrm{i}\sin \frac \theta 2 & \cos \frac \theta 2
\end{array}
\right) \left( 
\begin{array}{cc}
Z & X-\mathrm{i}Y \\ 
X+\mathrm{i}Y & -Z
\end{array}
\right) \left( 
\begin{array}{cc}
\cos \frac \theta 2 & -\mathrm{i}\sin \frac \theta 2 \\ 
-\mathrm{i}\sin \frac \theta 2 & \cos \frac \theta 2
\end{array}
\right) 
\]
\[
=\left( 
\begin{array}{cc}
Z\cos \theta -Y\sin \theta & X-\mathrm{i}\left( Y\cos \theta +Z\sin \theta
\right) \\ 
X+\mathrm{i}\left( Y\cos \theta +Z\sin \theta \right) & -Z\cos \theta +Y\sin
\theta
\end{array}
\right) \mbox{.} 
\]

Hence, if:

\[
U_{2,3}\left( \alpha \right) \stackrel{def}{=}\left[ 
\begin{array}{cccc}
\cos \alpha & \mathrm{i}\sin \alpha & 0 & 0 \\ 
\mathrm{i}\sin \alpha & \cos \alpha & 0 & 0 \\ 
0 & 0 & \cos \alpha & \mathrm{i}\sin \alpha \\ 
0 & 0 & \mathrm{i}\sin \alpha & \cos \alpha
\end{array}
\right] 
\]

and

\[
\widehat{M}^{\prime }\stackrel{def}{=} 
\begin{array}{c}
-M_{\zeta ,0}^{\prime }\gamma _\zeta ^{[0]}+M_{\zeta ,4}^{\prime }\zeta
^{[4]}- \\ 
-M_{\eta ,0}^{\prime }\gamma _\eta ^{[0]}-M_{\eta ,4}^{\prime }\eta ^{[4]}+
\\ 
+M_{\theta ,0}^{\prime }\gamma _\theta ^{[0]}+M_{\theta ,4}^{\prime }\theta
^{[4]}
\end{array}
\stackrel{def}{=}U_{2,3}^{\dagger }\left( \alpha \right) \widehat{M}%
U_{2,3}\left( \alpha \right) 
\]

then:

$M_{\zeta ,0}^{\prime }=M_{\zeta ,0}$,

$M_{\eta ,0}^{\prime }=M_{\eta ,0}\cos 2\alpha +M_{\theta ,0}\sin 2\alpha $,

$M_{\theta ,0}^{\prime }=M_{\theta ,0}\cos 2\alpha -M_{\eta ,0}\sin 2\alpha $%
,

$M_{\zeta ,4}^{\prime }=M_{\zeta ,4}$,

$M_{\eta ,4}^{\prime }=M_{\eta ,4}\cos 2\alpha +M_{\theta ,4}\sin 2\alpha $,

$M_{\theta ,4}^{\prime }=M_{\theta ,4}\cos 2\alpha -M_{\eta ,4}\sin 2\alpha $%
.

Therefore, matrix $U_{2,3}\left( \alpha \right) $ makes an oscillation
between green and blue quarks colors.

Let us consider equation (\ref{ham0}) under transformation $U_{2,3}\left(
\alpha \right) $ with $\alpha $ is an arbitrary real function of time-space
variables ($\alpha =\alpha \left( t,x_1,x_2,x_3\right) $):

\[
U_{2,3}^{\dagger }\left( \alpha \right) \left( \frac 1{\mathrm{c}}\partial
_t+\mathrm{i}\Theta _0+\mathrm{i}\Upsilon _0\gamma ^{\left[ 5\right]
}\right) U_{2,3}\left( \alpha \right) \varphi = 
\]

\[
=U_{2,3}^{\dagger }\left( \alpha \right) \left( 
\begin{array}{c}
\beta ^{\left[ 1\right] }\left( \partial _1+\mathrm{i}\Theta _1+\mathrm{i}%
\Upsilon _1\gamma ^{\left[ 5\right] }\right) + \\ 
+\beta ^{\left[ 2\right] }\left( \partial _2+\mathrm{i}\Theta _2+\mathrm{i}%
\Upsilon _2\gamma ^{\left[ 5\right] }\right) + \\ 
+\beta ^{\left[ 3\right] }\left( \partial _3+\mathrm{i}\Theta _3+\mathrm{i}%
\Upsilon _3\gamma ^{\left[ 5\right] }\right) + \\ 
+\mathrm{i}M_0\gamma ^{\left[ 0\right] }+\mathrm{i}M_4\beta ^{\left[
4\right] }+\widehat{M}
\end{array}
\right) U_{2,3}\left( \alpha \right) \varphi \mbox{.} 
\]

That is:

\[
\left( 
\begin{array}{c}
\frac 1{\mathrm{c}}U_{2,3}^{\dagger }\left( \alpha \right) \partial _t\left(
U_{2,3}\left( \alpha \right) \varphi \right) \\ 
+\mathrm{i}\Theta _0U_{2,3}^{\dagger }\left( \alpha \right) \left(
U_{2,3}\left( \alpha \right) \varphi \right) \\ 
+\mathrm{i}\Upsilon _0U_{2,3}^{\dagger }\left( \alpha \right) \gamma
^{\left[ 5\right] }\left( U_{2,3}\left( \alpha \right) \varphi \right)
\end{array}
\right) = 
\]

\[
=U_{2,3}^{\dagger }\left( \alpha \right) \left( 
\begin{array}{c}
\beta ^{\left[ 1\right] }\left( \partial _1+\mathrm{i}\Theta _1+\mathrm{i}%
\Upsilon _1\gamma ^{\left[ 5\right] }\right) + \\ 
+\beta ^{\left[ 2\right] }\left( \partial _2+\mathrm{i}\Theta _2+\mathrm{i}%
\Upsilon _2\gamma ^{\left[ 5\right] }\right) + \\ 
+\beta ^{\left[ 3\right] }\left( \partial _3+\mathrm{i}\Theta _3+\mathrm{i}%
\Upsilon _3\gamma ^{\left[ 5\right] }\right) + \\ 
+\mathrm{i}M_0\gamma ^{\left[ 0\right] }+\mathrm{i}M_4\beta ^{\left[
4\right] }+\widehat{M}
\end{array}
\right) U_{2,3}\left( \alpha \right) \varphi \mbox{.} 
\]

Because

\[
\begin{array}{c}
U_{2,3}^{\dagger }\left( \alpha \right) U_{2,3}\left( \alpha \right) =1_4%
\mbox{,} \\ 
U_{2,3}^{\dagger }\left( \alpha \right) \gamma ^{\left[ 5\right]
}U_{2,3}\left( \alpha \right) =\gamma ^{\left[ 5\right] }
\end{array}
\]

then

\[
\left( 
\begin{array}{c}
\frac 1{\mathrm{c}}\left( \left( U_{2,3}^{\dagger }\left( \alpha \right)
\partial _tU_{2,3}\left( \alpha \right) \right) \varphi +U_{2,3}^{\dagger
}\left( \alpha \right) U_{2,3}\left( \alpha \right) \partial _t\varphi
\right) \\ 
+\mathrm{i}\Theta _0\varphi +\mathrm{i}\Upsilon _0\gamma ^{\left[ 5\right]
}\varphi
\end{array}
\right) = 
\]

\[
=\left( 
\begin{array}{c}
U_{2,3}^{\dagger }\left( \alpha \right) \beta ^{\left[ 1\right] }\left( 
\begin{array}{c}
\left( \partial _1U_{2,3}\left( \alpha \right) \right) \varphi
+U_{2,3}\left( \alpha \right) \partial _1\varphi \\ 
+\mathrm{i}\Theta _1\left( U_{2,3}\left( \alpha \right) \varphi \right) +%
\mathrm{i}\Upsilon _1\gamma ^{\left[ 5\right] }\left( U_{2,3}\left( \alpha
\right) \varphi \right)
\end{array}
\right) + \\ 
+U_{2,3}^{\dagger }\left( \alpha \right) \beta ^{\left[ 2\right] }\left( 
\begin{array}{c}
\left( \partial _2U_{2,3}\left( \alpha \right) \right) \varphi
+U_{2,3}\left( \alpha \right) \partial _2\varphi \\ 
+\mathrm{i}\Theta _2\left( U_{2,3}\left( \alpha \right) \varphi \right) +%
\mathrm{i}\Upsilon _2\gamma ^{\left[ 5\right] }\left( U_{2,3}\left( \alpha
\right) \varphi \right)
\end{array}
\right) + \\ 
+U_{2,3}^{\dagger }\left( \alpha \right) \beta ^{\left[ 3\right] }\left( 
\begin{array}{c}
\left( \partial _3U_{2,3}\left( \alpha \right) \right) \varphi
+U_{2,3}\left( \alpha \right) \partial _3\varphi \\ 
+\mathrm{i}\Theta _3\left( U_{2,3}\left( \alpha \right) \varphi \right) +%
\mathrm{i}\Upsilon _3\gamma ^{\left[ 5\right] }\left( U_{2,3}\left( \alpha
\right) \varphi \right)
\end{array}
\right) + \\ 
+\mathrm{i}M_0U_{2,3}^{\dagger }\left( \alpha \right) \gamma ^{\left[
0\right] }\left( U_{2,3}\left( \alpha \right) \varphi \right) \\ 
+\mathrm{i}M_4U_{2,3}^{\dagger }\left( \alpha \right) \beta ^{\left[
4\right] }\left( U_{2,3}\left( \alpha \right) \varphi \right) \\ 
+U_{2,3}^{\dagger }\left( \alpha \right) \widehat{M}\left( U_{2,3}\left(
\alpha \right) \varphi \right)
\end{array}
\right) \mbox{.} 
\]

Since

\[
\begin{array}{c}
U_{2,3}^{\dagger }\left( \alpha \right) \gamma ^{\left[ 0\right]
}U_{2,3}\left( \alpha \right) =\gamma ^{\left[ 0\right] }\mbox{,} \\ 
U_{2,3}^{\dagger }\left( \alpha \right) \beta ^{\left[ 4\right]
}U_{2,3}\left( \alpha \right) =\beta ^{\left[ 4\right] }
\end{array}
\]

then

\[
\left( 
\begin{array}{c}
\frac 1{\mathrm{c}}\left( \left( U_{2,3}^{\dagger }\left( \alpha \right)
\partial _tU_{2,3}\left( \alpha \right) \right) \varphi +U_{2,3}^{\dagger
}\left( \alpha \right) U_{2,3}\left( \alpha \right) \partial _t\varphi
\right) \\ 
+\mathrm{i}\Theta _0\varphi +\mathrm{i}\Upsilon _0\gamma ^{\left[ 5\right]
}\varphi
\end{array}
\right) = 
\]

\[
=\left( 
\begin{array}{c}
U_{2,3}^{\dagger }\left( \alpha \right) \beta ^{\left[ 1\right] }\left( 
\begin{array}{c}
\left( \partial _1U_{2,3}\left( \alpha \right) \right) \varphi
+U_{2,3}\left( \alpha \right) \partial _1\varphi \\ 
+\mathrm{i}\Theta _1\left( U_{2,3}\left( \alpha \right) \varphi \right) +%
\mathrm{i}\Upsilon _1\gamma ^{\left[ 5\right] }\left( U_{2,3}\left( \alpha
\right) \varphi \right)
\end{array}
\right) + \\ 
+U_{2,3}^{\dagger }\left( \alpha \right) \beta ^{\left[ 2\right] }\left( 
\begin{array}{c}
\left( \partial _2U_{2,3}\left( \alpha \right) \right) \varphi
+U_{2,3}\left( \alpha \right) \partial _2\varphi \\ 
+\mathrm{i}\Theta _2\left( U_{2,3}\left( \alpha \right) \varphi \right) +%
\mathrm{i}\Upsilon _2\gamma ^{\left[ 5\right] }\left( U_{2,3}\left( \alpha
\right) \varphi \right)
\end{array}
\right) + \\ 
+U_{2,3}^{\dagger }\left( \alpha \right) \beta ^{\left[ 3\right] }\left( 
\begin{array}{c}
\left( \partial _3U_{2,3}\left( \alpha \right) \right) \varphi
+U_{2,3}\left( \alpha \right) \partial _3\varphi \\ 
+\mathrm{i}\Theta _3\left( U_{2,3}\left( \alpha \right) \varphi \right) +%
\mathrm{i}\Upsilon _3\gamma ^{\left[ 5\right] }\left( U_{2,3}\left( \alpha
\right) \varphi \right)
\end{array}
\right) + \\ 
+\mathrm{i}M_0\gamma ^{\left[ 0\right] }\varphi +\mathrm{i}M_4\beta ^{\left[
4\right] }\varphi +\widehat{M}^{\prime }\varphi
\end{array}
\right) \mbox{.} 
\]

Since

\[
U_{2,3}^{\dagger }\left( \alpha \right) \beta ^{\left[ 1\right] }=\beta
^{\left[ 1\right] }U_{2,3}^{\dagger }\left( \alpha \right) 
\]

then

\[
\left( U_{2,3}^{\dagger }\left( \alpha \right) \frac 1{\mathrm{c}}\partial
_tU_{2,3}\left( \alpha \right) +\frac 1{\mathrm{c}}\partial _t+\mathrm{i}%
\Theta _0+\mathrm{i}\Upsilon _0\gamma ^{\left[ 5\right] }\right) \varphi = 
\]

\[
=\left( 
\begin{array}{c}
\beta ^{\left[ 1\right] }U_{2,3}^{\dagger }\left( \alpha \right) \left( 
\begin{array}{c}
\left( \partial _1U_{2,3}\left( \alpha \right) \right) +U_{2,3}\left( \alpha
\right) \partial _1 \\ 
+\mathrm{i}\Theta _1U_{2,3}\left( \alpha \right) +\mathrm{i}\Upsilon
_1\gamma ^{\left[ 5\right] }U_{2,3}\left( \alpha \right)
\end{array}
\right) + \\ 
+U_{2,3}^{\dagger }\left( \alpha \right) \beta ^{\left[ 2\right] }\left( 
\begin{array}{c}
\left( \partial _2U_{2,3}\left( \alpha \right) \right) +U_{2,3}\left( \alpha
\right) \partial _2 \\ 
+\mathrm{i}\Theta _2U_{2,3}\left( \alpha \right) +\mathrm{i}\Upsilon
_2\gamma ^{\left[ 5\right] }U_{2,3}\left( \alpha \right)
\end{array}
\right) + \\ 
+U_{2,3}^{\dagger }\left( \alpha \right) \beta ^{\left[ 3\right] }\left( 
\begin{array}{c}
\left( \partial _3U_{2,3}\left( \alpha \right) \right) +U_{2,3}\left( \alpha
\right) \partial _3 \\ 
+\mathrm{i}\Theta _3U_{2,3}\left( \alpha \right) +\mathrm{i}\Upsilon
_3\gamma ^{\left[ 5\right] }U_{2,3}\left( \alpha \right)
\end{array}
\right) + \\ 
+\mathrm{i}M_0\gamma ^{\left[ 0\right] }+\mathrm{i}M_4\beta ^{\left[
4\right] }+\widehat{M}^{\prime }
\end{array}
\right) \varphi \mbox{.} 
\]

Because

\[
\begin{array}{c}
U_{2,3}^{\dagger }\left( \alpha \right) \beta ^{\left[ 2\right] }=\left(
\beta ^{\left[ 2\right] }\cos 2\alpha +\beta ^{\left[ 3\right] }\sin 2\alpha
\right) U_{2,3}^{\dagger }\left( \alpha \right) \mbox{,} \\ 
U_{2,3}^{\dagger }\left( \alpha \right) \beta ^{\left[ 3\right] }=\left(
\beta ^{\left[ 3\right] }\cos 2\alpha -\beta ^{\left[ 2\right] }\sin 2\alpha
\right) U_{2,3}^{\dagger }\left( \alpha \right)
\end{array}
\]

then

\[
\left( U_{2,3}^{\dagger }\left( \alpha \right) \frac 1{\mathrm{c}}\partial
_tU_{2,3}\left( \alpha \right) +\frac 1{\mathrm{c}}\partial _t+\mathrm{i}%
\Theta _0+\mathrm{i}\Upsilon _0\gamma ^{\left[ 5\right] }\right) \varphi = 
\]

\[
=\left( 
\begin{array}{c}
\beta ^{\left[ 1\right] }\left( 
\begin{array}{c}
U_{2,3}^{\dagger }\left( \alpha \right) \partial _1U_{2,3}\left( \alpha
\right) +U_{2,3}^{\dagger }\left( \alpha \right) U_{2,3}\left( \alpha
\right) \partial _1 \\ 
+\mathrm{i}\Theta _1U_{2,3}^{\dagger }\left( \alpha \right) U_{2,3}\left(
\alpha \right) +\mathrm{i}\Upsilon _1U_{2,3}^{\dagger }\left( \alpha \right)
\gamma ^{\left[ 5\right] }U_{2,3}\left( \alpha \right)
\end{array}
\right) + \\ 
+\left( \beta ^{\left[ 2\right] }\cos 2\alpha +\beta ^{\left[ 3\right] }\sin
2\alpha \right) \\ 
\left( 
\begin{array}{c}
U_{2,3}^{\dagger }\left( \alpha \right) \partial _2U_{2,3}\left( \alpha
\right) +U_{2,3}^{\dagger }\left( \alpha \right) U_{2,3}\left( \alpha
\right) \partial _2 \\ 
+\mathrm{i}\Theta _2U_{2,3}^{\dagger }\left( \alpha \right) U_{2,3}\left(
\alpha \right) +\mathrm{i}\Upsilon _2U_{2,3}^{\dagger }\left( \alpha \right)
\gamma ^{\left[ 5\right] }U_{2,3}\left( \alpha \right)
\end{array}
\right) + \\ 
+\left( \beta ^{\left[ 3\right] }\cos 2\alpha -\beta ^{\left[ 2\right] }\sin
2\alpha \right) \\ 
\left( 
\begin{array}{c}
U_{2,3}^{\dagger }\left( \alpha \right) \partial _3U_{2,3}\left( \alpha
\right) +U_{2,3}^{\dagger }\left( \alpha \right) U_{2,3}\left( \alpha
\right) \partial _3 \\ 
+\mathrm{i}\Theta _3U_{2,3}^{\dagger }\left( \alpha \right) U_{2,3}\left(
\alpha \right) +\mathrm{i}\Upsilon _3U_{2,3}^{\dagger }\left( \alpha \right)
\gamma ^{\left[ 5\right] }U_{2,3}\left( \alpha \right)
\end{array}
\right) + \\ 
+\mathrm{i}M_0\gamma ^{\left[ 0\right] }+\mathrm{i}M_4\beta ^{\left[
4\right] }+\widehat{M}^{\prime }
\end{array}
\right) \varphi \mbox{.} 
\]

Hence

\[
\left( \frac 1{\mathrm{c}}\partial _t+U_{2,3}^{\dagger }\left( \alpha
\right) \frac 1{\mathrm{c}}\partial _tU_{2,3}\left( \alpha \right) +\mathrm{i%
}\Theta _0+\mathrm{i}\Upsilon _0\gamma ^{\left[ 5\right] }\right) \varphi = 
\]

\begin{equation}
=\left( 
\begin{array}{c}
\beta ^{\left[ 1\right] }\left( \partial _1+U_{2,3}^{\dagger }\left( \alpha
\right) \partial _1U_{2,3}\left( \alpha \right) +\mathrm{i}\Theta _1+\mathrm{%
i}\Upsilon _1\gamma ^{\left[ 5\right] }\right) \\ 
+\beta ^{\left[ 2\right] }\left( 
\begin{array}{c}
\left( \cos 2\alpha \cdot \partial _2-\sin 2\alpha \cdot \partial _3\right)
\\ 
+U_{2,3}^{\dagger }\left( \alpha \right) \left( \cos 2\alpha \cdot \partial
_2-\sin 2\alpha \cdot \partial _3\right) U_{2,3}\left( \alpha \right) \\ 
+\mathrm{i}\left( \Theta _2\cos 2\alpha -\Theta _3\sin 2\alpha \right) \\ 
+\mathrm{i}\left( \Upsilon _2\gamma ^{\left[ 5\right] }\cos 2\alpha
-\Upsilon _3\gamma ^{\left[ 5\right] }\sin 2\alpha \right)
\end{array}
\right) \\ 
+\beta ^{\left[ 3\right] }\left( 
\begin{array}{c}
\left( \cos 2\alpha \cdot \partial _3+\sin 2\alpha \cdot \partial _2\right)
\\ 
+U_{2,3}^{\dagger }\left( \alpha \right) \left( \cos 2\alpha \cdot \partial
_3+\sin 2\alpha \cdot \partial _2\right) U_{2,3}\left( \alpha \right) \\ 
+\mathrm{i}\left( \Theta _2\sin 2\alpha +\Theta _3\cos 2\alpha \right) \\ 
+\mathrm{i}\left( \Upsilon _3\gamma ^{\left[ 5\right] }\cos 2\alpha
+\Upsilon _2\gamma ^{\left[ 5\right] }\sin 2\alpha \right)
\end{array}
\right) + \\ 
+\mathrm{i}M_0\gamma ^{\left[ 0\right] }+\mathrm{i}M_4\beta ^{\left[
4\right] }+\widehat{M}^{\prime }
\end{array}
\right) \varphi \mbox{.}  \label{ham5}
\end{equation}

Let $x_2^{\prime }$ and $x_3^{\prime }$ are elements of other coordinates
system such that:

\begin{eqnarray*}
\ &&\frac{\partial x_2}{\partial x_2^{\prime }}=\cos 2\alpha \mbox{,} \\
\ &&\frac{\partial x_3}{\partial x_2^{\prime }}=-\sin 2\alpha \mbox{,} \\
\ &&\frac{\partial x_2}{\partial x_3^{\prime }}=\sin 2\alpha \mbox{.} \\
\ &&\frac{\partial x_3}{\partial x_3^{\prime }}=\cos 2\alpha \mbox{,} \\
\ &&\frac{\partial x_0}{\partial x_2^{\prime }}=\frac{\partial x_1}{\partial
x_2^{\prime }}=\frac{\partial x_0}{\partial x_3^{\prime }}=\frac{\partial x_1%
}{\partial x_3^{\prime }}=0\mbox{.}
\end{eqnarray*}

Hence:

\begin{eqnarray*}
\partial _2^{\prime }\stackrel{def}{=}\frac \partial {\partial x_2^{\prime
}} &=&\frac \partial {\partial x_0}\frac{\partial x_0}{\partial x_2^{\prime }%
}+\frac \partial {\partial x_1}\frac{\partial x_1}{\partial x_2^{\prime }}%
+\frac \partial {\partial x_2}\frac{\partial x_2}{\partial x_2^{\prime }}%
+\frac \partial {\partial x_3}\frac{\partial x_3}{\partial x_2^{\prime }} \\
&=&\cos 2\alpha \cdot \frac \partial {\partial x_2}-\sin 2\alpha \cdot \frac
\partial {\partial x_3} \\
&=&\cos 2\alpha \cdot \partial _2-\sin 2\alpha \cdot \partial _3\mbox{,}
\end{eqnarray*}

\begin{eqnarray*}
\partial _3^{\prime }\stackrel{def}{=}\frac \partial {\partial x_3^{\prime
}} &=&\frac \partial {\partial x_0}\frac{\partial x_0}{\partial x_3^{\prime }%
}+\frac \partial {\partial x_1}\frac{\partial x_1}{\partial x_3^{\prime }}%
+\frac \partial {\partial x_2}\frac{\partial x_2}{\partial x_3^{\prime }}%
+\frac \partial {\partial x_3}\frac{\partial x_3}{\partial x_3^{\prime }} \\
\ &=&\cos 2\alpha \cdot \frac \partial {\partial x_3}+\sin 2\alpha \cdot
\frac \partial {\partial x_2} \\
\ &=&\cos 2\alpha \cdot \partial _3+\sin 2\alpha \cdot \partial _2.
\end{eqnarray*}

Therefore from (\ref{ham5}):

\[
\left( \frac 1{\mathrm{c}}\partial _t+U_{2,3}^{\dagger }\left( \alpha
\right) \frac 1{\mathrm{c}}\partial _tU_{2,3}\left( \alpha \right) +\mathrm{i%
}\Theta _0+\mathrm{i}\Upsilon _0\gamma ^{\left[ 5\right] }\right) \varphi = 
\]

\[
=\left( 
\begin{array}{c}
\beta ^{\left[ 1\right] }\left( \partial _1+U_{2,3}^{\dagger }\left( \alpha
\right) \partial _1U_{2,3}\left( \alpha \right) +\mathrm{i}\Theta _1+\mathrm{%
i}\Upsilon _1\gamma ^{\left[ 5\right] }\right) \\ 
+\beta ^{\left[ 2\right] }\left( \partial _2^{\prime }+U_{2,3}^{\dagger
}\left( \alpha \right) \partial _2^{\prime }U_{2,3}\left( \alpha \right) +%
\mathrm{i}\Theta _2^{\prime }+\mathrm{i}\Upsilon _2^{\prime }\gamma ^{\left[
5\right] }\right) \\ 
+\beta ^{\left[ 3\right] }\left( \partial _3^{\prime }+U_{2,3}^{\dagger
}\left( \alpha \right) \partial _3^{\prime }U_{2,3}\left( \alpha \right) +%
\mathrm{i}\Theta _3^{\prime }+\mathrm{i}\Upsilon _3^{\prime }\gamma ^{\left[
5\right] }\right) + \\ 
+\mathrm{i}M_0\gamma ^{\left[ 0\right] }+\mathrm{i}M_4\beta ^{\left[
4\right] }+\widehat{M}^{\prime }
\end{array}
\right) \varphi \mbox{.} 
\]

with

\[
\begin{array}{c}
\Theta _2^{\prime }\stackrel{def}{=}\Theta _2\cos 2\alpha -\Theta _3\sin
2\alpha \mbox{,} \\ 
\Theta _3^{\prime }\stackrel{def}{=}\Theta _2\sin 2\alpha +\Theta _3\cos
2\alpha \mbox{.} \\ 
\Upsilon _2^{\prime }\stackrel{def}{=}\Upsilon _2\cos 2\alpha -\Upsilon
_3\sin 2\alpha \mbox{,} \\ 
\Upsilon _3^{\prime }\stackrel{def}{=}\Upsilon _3\cos 2\alpha +\Upsilon
_2\sin 2\alpha \mbox{.}
\end{array}
\]

Therefore, the oscillation between blue and green quarks colors curves the
space of events in the $x_2$, $x_3$ directions.

Similarly that: matrix

\[
U_{1,3}\left( \vartheta \right) \stackrel{def}{=}\left[ 
\begin{array}{cccc}
\cos \vartheta & \sin \vartheta & 0 & 0 \\ 
-\sin \vartheta & \cos \vartheta & 0 & 0 \\ 
0 & 0 & \cos \vartheta & \sin \vartheta \\ 
0 & 0 & -\sin \vartheta & \cos \vartheta
\end{array}
\right] 
\]

with an arbitrary real function $\vartheta \left( t,x_1,x_2,x_3\right) $
describes the oscillation between blue and red quarks colors which curves
the space of events in the $x_1$, $x_3$ directions. And matrix

\[
U_{1,2}\left( \varsigma \right) \stackrel{def}{=}\left[ 
\begin{array}{cccc}
\cos \varsigma -\mathrm{i}\sin \varsigma & 0 & 0 & 0 \\ 
0 & \cos \varsigma +\mathrm{i}\sin \varsigma & 0 & 0 \\ 
0 & 0 & \cos \varsigma -\mathrm{i}\sin \varsigma & 0 \\ 
0 & 0 & 0 & \cos \varsigma +\mathrm{i}\sin \varsigma
\end{array}
\right] 
\]

with an arbitrary real function $\varsigma \left( t,x_1,x_2,x_3\right) $
describes the oscillation between green and red quarks colors which curves
the space of events in the $x_1$, $x_2$ directions.

Now, let

\[
U_{0,1}\left( \varpi \right) \stackrel{def}{=}\left[ 
\begin{array}{cccc}
\cosh \varpi & -\sinh \varpi & 0 & 0 \\ 
-\sinh \varpi & \cosh \varpi & 0 & 0 \\ 
0 & 0 & \cosh \varpi & \sinh \varpi \\ 
0 & 0 & \sinh \varpi & \cosh \varpi
\end{array}
\right] \mbox{.} 
\]

and

\[
\widehat{M}^{\prime \prime }\stackrel{def}{=} 
\begin{array}{c}
-M_{\zeta ,0}^{\prime \prime }\gamma _\zeta ^{[0]}+M_{\zeta ,4}^{\prime
\prime }\zeta ^{[4]}- \\ 
-M_{\eta ,0}^{\prime \prime }\gamma _\eta ^{[0]}-M_{\eta ,4}^{\prime \prime
}\eta ^{[4]}+ \\ 
+M_{\theta ,0}^{\prime \prime }\gamma _\theta ^{[0]}+M_{\theta ,4}^{\prime
\prime }\theta ^{[4]}
\end{array}
\stackrel{def}{=}U_{0,1}^{\dagger }\left( \varpi \right) \widehat{M}%
U_{0,1}\left( \varpi \right) 
\]

then:

$M_{\zeta ,0}^{\prime \prime }=M_{\zeta ,0}$,

$M_{\eta ,0}^{\prime \prime }=\left( M_{\eta ,0}\cosh 2\varpi -M_{\theta
,4}\sinh 2\varpi \right) $,

$M_{\theta ,0}^{\prime \prime }=M_{\theta ,0}\cosh 2\varpi +M_{\eta
,4}\sinh 2\varpi $,

$M_{\zeta ,4}^{\prime \prime }=M_{\zeta ,4}$

$M_{\eta ,4}^{\prime \prime }=M_{\eta ,4}\cosh 2\varpi +M_{\theta ,0}\sinh
2\varpi $,

$M_{\theta ,4}^{\prime \prime }=M_{\theta ,4}\cosh 2\varpi -M_{\eta
,0}\sinh 2\varpi $,

Therefore, matrix $U_{0,1}\left( \varpi \right) $ makes an oscillation
between green and blue quarks colors with an oscillation between up and down
quarks.

Let us consider equation (\ref{ham0}) under transformation $U_{0,1}\left(
\varpi \right) $ with $\varpi $ is an arbitrary real function of
time-space variables ($\varpi =\varpi \left( t,x_1,x_2,x_3\right) $):

\[
U_{0,1}^{\dagger }\left( \varpi \right) \left( \frac 1{\mathrm{c}}\partial
_t+\mathrm{i}\Theta _0+\mathrm{i}\Upsilon _0\gamma ^{\left[ 5\right]
}\right) U_{0,1}\left( \varpi \right) \varphi = 
\]

\[
=U_{0,1}^{\dagger }\left( \varpi \right) \left( 
\begin{array}{c}
\beta ^{\left[ 1\right] }\left( \partial _1+\mathrm{i}\Theta _1+\mathrm{i}%
\Upsilon _1\gamma ^{\left[ 5\right] }\right) + \\ 
+\beta ^{\left[ 2\right] }\left( \partial _2+\mathrm{i}\Theta _2+\mathrm{i}%
\Upsilon _2\gamma ^{\left[ 5\right] }\right) + \\ 
+\beta ^{\left[ 3\right] }\left( \partial _3+\mathrm{i}\Theta _3+\mathrm{i}%
\Upsilon _3\gamma ^{\left[ 5\right] }\right) + \\ 
+\mathrm{i}M_0\gamma ^{\left[ 0\right] }+\mathrm{i}M_4\beta ^{\left[
4\right] }+\widehat{M}
\end{array}
\right) U_{0,1}\left( \varpi \right) \varphi \mbox{.} 
\]

Hence,

\[
U_{0,1}^{\dagger }\left( \varpi \right) \left( 
\begin{array}{c}
\frac 1{\mathrm{c}}\partial _t\left( U_{0,1}\left( \varpi \right) \varphi
\right) \\ 
+\mathrm{i}\Theta _0\left( U_{0,1}\left( \varpi \right) \varphi \right) +%
\mathrm{i}\Upsilon _0\gamma ^{\left[ 5\right] }\left( U_{0,1}\left( \varpi
\right) \varphi \right)
\end{array}
\right) = 
\]

\[
=U_{0,1}^{\dagger }\left( \varpi \right) \left( 
\begin{array}{c}
\beta ^{\left[ 1\right] }\left( 
\begin{array}{c}
\partial _1\left( U_{0,1}\left( \varpi \right) \varphi \right) \\ 
+\mathrm{i}\Theta _1\left( U_{0,1}\left( \varpi \right) \varphi \right) +%
\mathrm{i}\Upsilon _1\gamma ^{\left[ 5\right] }\left( U_{0,1}\left( \varpi
\right) \varphi \right)
\end{array}
\right) + \\ 
+\beta ^{\left[ 2\right] }\left( 
\begin{array}{c}
\partial _2\left( U_{0,1}\left( \varpi \right) \varphi \right) \\ 
+\mathrm{i}\Theta _2\left( U_{0,1}\left( \varpi \right) \varphi \right) +%
\mathrm{i}\Upsilon _2\gamma ^{\left[ 5\right] }\left( U_{0,1}\left( \varpi
\right) \varphi \right)
\end{array}
\right) + \\ 
+\beta ^{\left[ 3\right] }\left( 
\begin{array}{c}
\partial _3\left( U_{0,1}\left( \varpi \right) \varphi \right) \\ 
+\mathrm{i}\Theta _3\left( U_{0,1}\left( \varpi \right) \varphi \right) +%
\mathrm{i}\Upsilon _3\gamma ^{\left[ 5\right] }\left( U_{0,1}\left( \varpi
\right) \varphi \right)
\end{array}
\right) + \\ 
+\mathrm{i}M_0\gamma ^{\left[ 0\right] }\left( U_{0,1}\left( \varpi
\right) \varphi \right) \\ 
+\mathrm{i}M_4\beta ^{\left[ 4\right] }\left( U_{0,1}\left( \varpi \right)
\varphi \right) +\widehat{M}\left( U_{0,1}\left( \varpi \right) \varphi
\right)
\end{array}
\right) \mbox{.} 
\]

\[
U_{0,1}^{\dagger }\left( \varpi \right) \left( 
\begin{array}{c}
\frac 1{\mathrm{c}}\left( \partial _tU_{0,1}\left( \varpi \right) \right)
\varphi +U_{0,1}\left( \varpi \right) \frac 1{\mathrm{c}}\partial _t\varphi
\\ 
+\mathrm{i}\Theta _0U_{0,1}\left( \varpi \right) \varphi +\mathrm{i}%
\Upsilon _0\gamma ^{\left[ 5\right] }U_{0,1}\left( \varpi \right) \varphi
\end{array}
\right) = 
\]

\[
=U_{0,1}^{\dagger }\left( \varpi \right) \left( 
\begin{array}{c}
\beta ^{\left[ 1\right] }\left( 
\begin{array}{c}
\left( \partial _1U_{0,1}\left( \varpi \right) \right) \varphi
+U_{0,1}\left( \varpi \right) \partial _1\varphi \\ 
+\mathrm{i}\Theta _1U_{0,1}\left( \varpi \right) \varphi +\mathrm{i}%
\Upsilon _1\gamma ^{\left[ 5\right] }U_{0,1}\left( \varpi \right) \varphi
\end{array}
\right) + \\ 
+\beta ^{\left[ 2\right] }\left( 
\begin{array}{c}
\left( \partial _2U_{0,1}\left( \varpi \right) \right) \varphi
+U_{0,1}\left( \varpi \right) \partial _2\varphi \\ 
+\mathrm{i}\Theta _2U_{0,1}\left( \varpi \right) \varphi +\mathrm{i}%
\Upsilon _2\gamma ^{\left[ 5\right] }U_{0,1}\left( \varpi \right) \varphi
\end{array}
\right) + \\ 
+\beta ^{\left[ 3\right] }\left( 
\begin{array}{c}
\left( \partial _3U_{0,1}\left( \varpi \right) \right) \varphi
+U_{0,1}\left( \varpi \right) \partial _3\varphi \\ 
+\mathrm{i}\Theta _3U_{0,1}\left( \varpi \right) \varphi +\mathrm{i}%
\Upsilon _3\gamma ^{\left[ 5\right] }U_{0,1}\left( \varpi \right) \varphi
\end{array}
\right) + \\ 
+\mathrm{i}M_0\gamma ^{\left[ 0\right] }U_{0,1}\left( \varpi \right)
\varphi \\ 
+\mathrm{i}M_4\beta ^{\left[ 4\right] }U_{0,1}\left( \varpi \right)
\varphi +\widehat{M}U_{0,1}\left( \varpi \right) \varphi
\end{array}
\right) \mbox{.} 
\]

Therefore:

\[
\left( 
\begin{array}{c}
U_{0,1}^{\dagger }\left( \varpi \right) \frac 1{\mathrm{c}}\left( \partial
_tU_{0,1}\left( \varpi \right) \right) +U_{0,1}^{\dagger }\left( \varpi
\right) U_{0,1}\left( \varpi \right) \frac 1{\mathrm{c}}\partial _t \\ 
+\mathrm{i}\Theta _0U_{0,1}^{\dagger }\left( \varpi \right) U_{0,1}\left(
\varpi \right) +\mathrm{i}\Upsilon _0U_{0,1}^{\dagger }\left( \varpi
\right) \gamma ^{\left[ 5\right] }U_{0,1}\left( \varpi \right)
\end{array}
\right) \varphi = 
\]

\[
=\left( 
\begin{array}{c}
U_{0,1}^{\dagger }\left( \varpi \right) \beta ^{\left[ 1\right] }\left( 
\begin{array}{c}
\left( \partial _1U_{0,1}\left( \varpi \right) \right) +U_{0,1}\left(
\varpi \right) \partial _1 \\ 
+\mathrm{i}\Theta _1U_{0,1}\left( \varpi \right) +\mathrm{i}\Upsilon
_1\gamma ^{\left[ 5\right] }U_{0,1}\left( \varpi \right)
\end{array}
\right) + \\ 
+U_{0,1}^{\dagger }\left( \varpi \right) \beta ^{\left[ 2\right] }\left( 
\begin{array}{c}
\left( \partial _2U_{0,1}\left( \varpi \right) \right) +U_{0,1}\left(
\varpi \right) \partial _2 \\ 
+\mathrm{i}\Theta _2U_{0,1}\left( \varpi \right) +\mathrm{i}\Upsilon
_2\gamma ^{\left[ 5\right] }U_{0,1}\left( \varpi \right)
\end{array}
\right) + \\ 
+U_{0,1}^{\dagger }\left( \varpi \right) \beta ^{\left[ 3\right] }\left( 
\begin{array}{c}
\left( \partial _3U_{0,1}\left( \varpi \right) \right) +U_{0,1}\left(
\varpi \right) \partial _3 \\ 
+\mathrm{i}\Theta _3U_{0,1}\left( \varpi \right) +\mathrm{i}\Upsilon
_3\gamma ^{\left[ 5\right] }U_{0,1}\left( \varpi \right)
\end{array}
\right) + \\ 
+\mathrm{i}M_0U_{0,1}^{\dagger }\left( \varpi \right) \gamma ^{\left[
0\right] }U_{0,1}\left( \varpi \right) \\ 
+\mathrm{i}M_4U_{0,1}^{\dagger }\left( \varpi \right) \beta ^{\left[
4\right] }U_{0,1}\left( \varpi \right) +U_{0,1}^{\dagger }\left( \varpi
\right) \widehat{M}U_{0,1}\left( \varpi \right)
\end{array}
\right) \varphi \mbox{.} 
\]

Since:

\begin{eqnarray*}
U_{0,1}^{\dagger }\left( \varpi \right) U_{0,1}\left( \varpi \right)
&=&\left( \cosh 2\varpi -\beta ^{\left[ 1\right] }\sinh 2\varpi \right) %
\mbox{,} \\
U_{0,1}^{\dagger }\left( \varpi \right) &=&\left( \cosh 2\varpi +\beta
^{\left[ 1\right] }\sinh 2\varpi \right) U_{0,1}^{-1}\left( \varpi
\right) \mbox{,} \\
U_{0,1}^{\dagger }\left( \varpi \right) \beta ^{\left[ 1\right] }
&=&\left( \beta ^{\left[ 1\right] }\cosh 2\varpi -\sinh 2\varpi \right)
U_{0,1}^{-1}\left( \varpi \right) \mbox{,} \\
U_{0,1}^{\dagger }\left( \varpi \right) \beta ^{\left[ 2\right] } &=&\beta
^{\left[ 2\right] }U_{0,1}^{-1}\left( \varpi \right) \mbox{,} \\
U_{0,1}^{\dagger }\left( \varpi \right) \beta ^{\left[ 3\right] } &=&\beta
^{\left[ 3\right] }U_{0,1}^{-1}\left( \varpi \right) \mbox{,} \\
U_{0,1}^{\dagger }\left( \varpi \right) \gamma ^{\left[ 0\right]
}U_{0,1}\left( \varpi \right) &=&\gamma ^{\left[ 0\right] }\mbox{,} \\
U_{0,1}^{\dagger }\left( \varpi \right) \beta ^{\left[ 4\right]
}U_{0,1}\left( \varpi \right) &=&\beta ^{\left[ 4\right] }
\end{eqnarray*}

then

\[
\left( 
\begin{array}{c}
\left( \cosh 2\varpi +\beta ^{\left[ 1\right] }\sinh 2\varpi \right)
U_{0,1}^{-1}\left( \varpi \right) \frac 1{\mathrm{c}}\left( \partial
_tU_{0,1}\left( \varpi \right) \right) \\ 
+\left( \cosh 2\varpi -\beta ^{\left[ 1\right] }\sinh 2\varpi \right)
\frac 1{\mathrm{c}}\partial _t \\ 
+\mathrm{i}\Theta _0\left( \cosh 2\varpi -\beta ^{\left[ 1\right] }\sinh
2\varpi \right) \\ 
+\mathrm{i}\Upsilon _0U_{0,1}^{\dagger }\left( \varpi \right) \gamma
^{\left[ 5\right] }U_{0,1}\left( \varpi \right)
\end{array}
\right) \varphi = 
\]

\[
=\left( 
\begin{array}{c}
\left( \beta ^{\left[ 1\right] }\cosh 2\varpi -\sinh 2\varpi \right)
\cdot \\ 
\cdot U_{0,1}^{-1}\left( \varpi \right) \left( 
\begin{array}{c}
\left( \partial _1U_{0,1}\left( \varpi \right) \right) +U_{0,1}\left(
\varpi \right) \partial _1 \\ 
+\mathrm{i}\Theta _1U_{0,1}\left( \varpi \right) +\mathrm{i}\Upsilon
_1\gamma ^{\left[ 5\right] }U_{0,1}\left( \varpi \right)
\end{array}
\right) + \\ 
+\beta ^{\left[ 2\right] }U_{0,1}^{-1}\left( \varpi \right) \left( 
\begin{array}{c}
\left( \partial _2U_{0,1}\left( \varpi \right) \right) +U_{0,1}\left(
\varpi \right) \partial _2 \\ 
+\mathrm{i}\Theta _2U_{0,1}\left( \varpi \right) +\mathrm{i}\Upsilon
_2\gamma ^{\left[ 5\right] }U_{0,1}\left( \varpi \right)
\end{array}
\right) + \\ 
+\beta ^{\left[ 3\right] }U_{0,1}^{-1}\left( \varpi \right) \left( 
\begin{array}{c}
\left( \partial _3U_{0,1}\left( \varpi \right) \right) +U_{0,1}\left(
\varpi \right) \partial _3 \\ 
+\mathrm{i}\Theta _3U_{0,1}\left( \varpi \right) +\mathrm{i}\Upsilon
_3\gamma ^{\left[ 5\right] }U_{0,1}\left( \varpi \right)
\end{array}
\right) + \\ 
+\mathrm{i}M_0\gamma ^{\left[ 0\right] }+\mathrm{i}M_4\beta ^{\left[
4\right] }+\widehat{M}^{\prime \prime }
\end{array}
\right) \varphi \mbox{.} 
\]

Hence:

\[
\left( 
\begin{array}{c}
\left( \cosh 2\varpi +\beta ^{\left[ 1\right] }\sinh 2\varpi \right)
U_{0,1}^{-1}\left( \varpi \right) \frac 1{\mathrm{c}}\left( \partial
_tU_{0,1}\left( \varpi \right) \right) \\ 
+\left( \cosh 2\varpi -\beta ^{\left[ 1\right] }\sinh 2\varpi \right)
\frac 1{\mathrm{c}}\partial _t \\ 
+\mathrm{i}\Theta _0\left( \cosh 2\varpi -\beta ^{\left[ 1\right] }\sinh
2\varpi \right) \\ 
+\mathrm{i}\Upsilon _0U_{0,1}^{\dagger }\left( \varpi \right) \gamma
^{\left[ 5\right] }U_{0,1}\left( \varpi \right)
\end{array}
\right) \varphi = 
\]

\[
=\left( 
\begin{array}{c}
\left( \beta ^{\left[ 1\right] }\cosh 2\varpi -\sinh 2\varpi \right)
\cdot \\ 
\cdot \left( 
\begin{array}{c}
U_{0,1}^{-1}\left( \varpi \right) \left( \partial _1U_{0,1}\left( \varpi
\right) \right) +U_{0,1}^{-1}\left( \varpi \right) U_{0,1}\left( \varpi
\right) \partial _1 \\ 
+\mathrm{i}\Theta _1U_{0,1}^{-1}\left( \varpi \right) U_{0,1}\left(
\varpi \right) +\mathrm{i}\Upsilon _1U_{0,1}^{-1}\left( \varpi \right)
\gamma ^{\left[ 5\right] }U_{0,1}\left( \varpi \right)
\end{array}
\right) + \\ 
+\beta ^{\left[ 2\right] }\left( 
\begin{array}{c}
U_{0,1}^{-1}\left( \varpi \right) \left( \partial _2U_{0,1}\left( \varpi
\right) \right) +U_{0,1}^{-1}\left( \varpi \right) U_{0,1}\left( \varpi
\right) \partial _2 \\ 
+\mathrm{i}\Theta _2U_{0,1}^{-1}\left( \varpi \right) U_{0,1}\left(
\varpi \right) +\mathrm{i}\Upsilon _2U_{0,1}^{-1}\left( \varpi \right)
\gamma ^{\left[ 5\right] }U_{0,1}\left( \varpi \right)
\end{array}
\right) + \\ 
+\beta ^{\left[ 3\right] }\left( 
\begin{array}{c}
U_{0,1}^{-1}\left( \varpi \right) \left( \partial _3U_{0,1}\left( \varpi
\right) \right) +U_{0,1}^{-1}\left( \varpi \right) U_{0,1}\left( \varpi
\right) \partial _3 \\ 
+\mathrm{i}\Theta _3U_{0,1}^{-1}\left( \varpi \right) U_{0,1}\left(
\varpi \right) +\mathrm{i}\Upsilon _3U_{0,1}^{-1}\left( \varpi \right)
\gamma ^{\left[ 5\right] }U_{0,1}\left( \varpi \right)
\end{array}
\right) + \\ 
+\mathrm{i}M_0\gamma ^{\left[ 0\right] }+\mathrm{i}M_4\beta ^{\left[
4\right] }+\widehat{M}^{\prime \prime }
\end{array}
\right) \varphi \mbox{.} 
\]

Because

\begin{eqnarray*}
U_{0,1}^{-1}\left( \varpi \right) U_{0,1}\left( \varpi \right) &=&1_4%
\mbox{,} \\
U_{0,1}^{-1}\left( \varpi \right) \gamma ^{\left[ 5\right] }U_{0,1}\left(
\varpi \right) &=&\gamma ^{\left[ 5\right] }\mbox{,} \\
U_{0,1}^{\dagger }\left( \varpi \right) \gamma ^{\left[ 5\right]
}U_{0,1}\left( \varpi \right) &=&\gamma ^{\left[ 5\right] }\left( \cosh
2\varpi -\beta ^{\left[ 1\right] }\sinh 2\varpi \right)
\end{eqnarray*}

then

\[
\left( 
\begin{array}{c}
\left( \cosh 2\varpi +\beta ^{\left[ 1\right] }\sinh 2\varpi \right)
U_{0,1}^{-1}\left( \varpi \right) \frac 1{\mathrm{c}}\partial
_tU_{0,1}\left( \varpi \right) \\ 
+\left( \cosh 2\varpi -\beta ^{\left[ 1\right] }\sinh 2\varpi \right)
\frac 1{\mathrm{c}}\partial _t \\ 
+\mathrm{i}\Theta _0\left( \cosh 2\varpi -\beta ^{\left[ 1\right] }\sinh
2\varpi \right) \\ 
+\mathrm{i}\Upsilon _0\gamma ^{\left[ 5\right] }\left( \cosh 2\varpi
-\beta ^{\left[ 1\right] }\sinh 2\varpi \right)
\end{array}
\right) \varphi = 
\]

\[
=\left( 
\begin{array}{c}
\left( \beta ^{\left[ 1\right] }\cosh 2\varpi -\sinh 2\varpi \right)
\cdot \\ 
\cdot \left( \partial _1+U_{0,1}^{-1}\left( \varpi \right) \left( \partial
_1U_{0,1}\left( \varpi \right) \right) +\mathrm{i}\Theta _1+\mathrm{i}%
\Upsilon _1\gamma ^{\left[ 5\right] }\right) + \\ 
+\beta ^{\left[ 2\right] }\left( \partial _2+U_{0,1}^{-1}\left( \varpi
\right) \left( \partial _2U_{0,1}\left( \varpi \right) \right) +\mathrm{i}%
\Theta _2+\mathrm{i}\Upsilon _2\gamma ^{\left[ 5\right] }\right) + \\ 
+\beta ^{\left[ 3\right] }\left( \partial _3+U_{0,1}^{-1}\left( \varpi
\right) \left( \partial _3U_{0,1}\left( \varpi \right) \right) +\mathrm{i}%
\Theta _3+\mathrm{i}\Upsilon _3\gamma ^{\left[ 5\right] }\right) + \\ 
+\mathrm{i}M_0\gamma ^{\left[ 0\right] }+\mathrm{i}M_4\beta ^{\left[
4\right] }+\widehat{M}^{\prime \prime }
\end{array}
\right) \varphi \mbox{.} 
\]

Hence:

\[
\left( 
\begin{array}{c}
\cosh 2\varpi \cdot U_{0,1}^{-1}\left( \varpi \right) \frac 1{\mathrm{c}%
}\partial _tU_{0,1}\left( \varpi \right) +\cosh 2\varpi \cdot \frac 1{%
\mathrm{c}}\partial _t+\sinh 2\varpi \cdot \partial _1 \\ 
+\mathrm{i}\Theta _0\cosh 2\varpi +\mathrm{i}\Upsilon _0\gamma ^{\left[
5\right] }\cosh 2\varpi \\ 
+\sinh 2\varpi \cdot U_{0,1}^{-1}\left( \varpi \right) \left( \partial
_1U_{0,1}\left( \varpi \right) \right) \\ 
+\mathrm{i}\Theta _1\sinh 2\varpi \\ 
+\mathrm{i}\sinh 2\varpi \cdot \Upsilon _1\gamma ^{\left[ 5\right] } \\ 
+\beta ^{\left[ 1\right] }\sinh 2\varpi \cdot U_{0,1}^{-1}\left( \varpi
\right) \frac 1{\mathrm{c}}\partial _tU_{0,1}\left( \varpi \right) -\beta
^{\left[ 1\right] }\sinh 2\varpi \cdot \frac 1{\mathrm{c}}\partial _t \\ 
-\mathrm{i}\beta ^{\left[ 1\right] }\Theta _0\sinh 2\varpi -\mathrm{i}%
\beta ^{\left[ 1\right] }\Upsilon _0\gamma ^{\left[ 5\right] }\sinh
2\varpi -\beta ^{\left[ 1\right] }\cosh 2\varpi \cdot \partial _1 \\ 
-\beta ^{\left[ 1\right] }\cosh 2\varpi \cdot U_{0,1}^{-1}\left( \varpi
\right) \left( \partial _1U_{0,1}\left( \varpi \right) \right) \\ 
-\mathrm{i}\beta ^{\left[ 1\right] }\Theta _1\cosh 2\varpi \\ 
-\mathrm{i}\beta ^{\left[ 1\right] }\cosh 2\varpi \cdot \Upsilon _1\gamma
^{\left[ 5\right] } \\ 
-\beta ^{\left[ 2\right] }\left( \partial _2+U_{0,1}^{-1}\left( \varpi
\right) \left( \partial _2U_{0,1}\left( \varpi \right) \right) +\mathrm{i}%
\Theta _2+\mathrm{i}\Upsilon _2\gamma ^{\left[ 5\right] }\right) \\ 
-\beta ^{\left[ 3\right] }\left( \partial _3+U_{0,1}^{-1}\left( \varpi
\right) \left( \partial _3U_{0,1}\left( \varpi \right) \right) +\mathrm{i}%
\Theta _3+\mathrm{i}\Upsilon _3\gamma ^{\left[ 5\right] }\right) \\ 
-\mathrm{i}M_0\gamma ^{\left[ 0\right] }-\mathrm{i}M_4\beta ^{\left[
4\right] }-\widehat{M}^{\prime \prime }
\end{array}
\right) \varphi =0\mbox{.} 
\]

Therefore,

\begin{equation}
\left( 
\begin{array}{c}
U_{0,1}^{-1}\left( \varpi \right) \left( \cosh 2\varpi \cdot \frac 1{%
\mathrm{c}}\partial _t+\sinh 2\varpi \cdot \partial _1\right)
U_{0,1}\left( \varpi \right) \\ 
+\left( \cosh 2\varpi \cdot \frac 1{\mathrm{c}}\partial _t+\sinh 2\varpi
\cdot \partial _1\right) \\ 
+\mathrm{i}\left( \Theta _0\cosh 2\varpi +\Theta _1\sinh 2\varpi \right)
\\ 
+\mathrm{i}\left( \Upsilon _0\cosh 2\varpi +\sinh 2\varpi \cdot \Upsilon
_1\right) \gamma ^{\left[ 5\right] } \\ 
-\beta ^{\left[ 1\right] }\left( 
\begin{array}{c}
U_{0,1}^{-1}\left( \varpi \right) \left( \cosh 2\varpi \cdot \partial
_1+\sinh 2\varpi \cdot \frac 1{\mathrm{c}}\partial _t\right) U_{0,1}\left(
\varpi \right) \\ 
+\left( \cosh 2\varpi \cdot \partial _1+\sinh 2\varpi \cdot \frac 1{%
\mathrm{c}}\partial _t\right) \\ 
+\mathrm{i}\left( \Theta _1\cosh 2\varpi +\Theta _0\sinh 2\varpi \right)
\\ 
+\mathrm{i}\left( \Upsilon _1\cosh 2\varpi +\Upsilon _0\sinh 2\varpi
\right) \gamma ^{\left[ 5\right] }
\end{array}
\right) \\ 
-\beta ^{\left[ 2\right] }\left( \partial _2+U_{0,1}^{-1}\left( \varpi
\right) \left( \partial _2U_{0,1}\left( \varpi \right) \right) +\mathrm{i}%
\Theta _2+\mathrm{i}\Upsilon _2\gamma ^{\left[ 5\right] }\right) \\ 
-\beta ^{\left[ 3\right] }\left( \partial _3+U_{0,1}^{-1}\left( \varpi
\right) \left( \partial _3U_{0,1}\left( \varpi \right) \right) +\mathrm{i}%
\Theta _3+\mathrm{i}\Upsilon _3\gamma ^{\left[ 5\right] }\right) \\ 
-\mathrm{i}M_0\gamma ^{\left[ 0\right] }-\mathrm{i}M_4\beta ^{\left[
4\right] }-\widehat{M}^{\prime \prime }
\end{array}
\right) \varphi =0  \label{ham6}
\end{equation}

Let $t^{\prime }$ and $x_1^{\prime }$ are elements of other coordinates
system such that:

\begin{eqnarray}
\  &&\frac{\partial x_1}{\partial x_1^{\prime }}=\cosh 2\varpi \mbox{,}
\label{grg} \\
\  &&\frac{\partial t}{\partial x_1^{\prime }}=\frac 1{\mathrm{c}}\sinh
2\varpi \mbox{,}  \nonumber \\
\  &&\frac{\partial x_1}{\partial t^{\prime }}=\mathrm{c}\sinh 2\varpi %
\mbox{.}  \nonumber \\
\  &&\frac{\partial t}{\partial t^{\prime }}=\cosh 2\varpi \mbox{,} 
\nonumber \\
\  &&\frac{\partial x_2}{\partial t^{\prime }}=\frac{\partial x_3}{\partial
t^{\prime }}=\frac{\partial x_2}{\partial x_1^{\prime }}=\frac{\partial x_3}{%
\partial x_1^{\prime }}=0\mbox{.}  \nonumber
\end{eqnarray}

Hence:

\begin{eqnarray*}
\partial _t^{\prime }\stackrel{def}{=}\frac \partial {\partial t^{\prime }}
&=&\frac \partial {\partial t}\frac{\partial t}{\partial t^{\prime }}+\frac
\partial {\partial x_1}\frac{\partial x_1}{\partial t^{\prime }}+\frac
\partial {\partial x_2}\frac{\partial x_2}{\partial t^{\prime }}+\frac
\partial {\partial x_3}\frac{\partial x_3}{\partial t^{\prime }} \\
\ &=&\cosh 2\varpi \cdot \frac \partial {\partial t}+\mathrm{c}\sinh
2\varpi \cdot \frac \partial {\partial x_1} \\
\ &=&\cosh 2\varpi \cdot \partial _t+\mathrm{c}\sinh 2\varpi \cdot
\partial _1\mbox{,}
\end{eqnarray*}

That is:

\[
\frac 1{\mathrm{c}}\partial _t^{\prime }=\frac 1{\mathrm{c}}\cosh 2\varpi
\cdot \partial _t+\sinh 2\varpi \cdot \partial _1\mbox{.} 
\]

And

\begin{eqnarray*}
\partial _1^{\prime }\stackrel{def}{=}\frac \partial {\partial x_1^{\prime
}} &=&\frac \partial {\partial t}\frac{\partial t}{\partial x_1^{\prime }}%
+\frac \partial {\partial x_1}\frac{\partial x_1}{\partial x_1^{\prime }}%
+\frac \partial {\partial x_2}\frac{\partial x_2}{\partial x_1^{\prime }}%
+\frac \partial {\partial x_3}\frac{\partial x_3}{\partial x_1^{\prime }} \\
\ &=&\cosh 2\varpi \cdot \frac \partial {\partial x_1}+\sinh 2\varpi
\cdot \frac 1{\mathrm{c}}\frac \partial {\partial t} \\
\ &=&\cosh 2\varpi \cdot \partial _1+\sinh 2\varpi \cdot \frac 1{\mathrm{%
c}}\partial _t.
\end{eqnarray*}

Therefore from (\ref{ham6}):

\[
\left( 
\begin{array}{c}
\beta ^{\left[ 0\right] }\left( \frac 1{\mathrm{c}}\partial _t^{\prime
}+U_{0,1}^{-1}\left( \varpi \right) \frac 1{\mathrm{c}}\partial _t^{\prime
}U_{0,1}\left( \varpi \right) +\mathrm{i}\Theta _0^{\prime \prime }+%
\mathrm{i}\Upsilon _0^{\prime \prime }\gamma ^{\left[ 5\right] }\right) \\ 
+\beta ^{\left[ 1\right] }\left( \partial _1^{\prime }+U_{0,1}^{-1}\left(
\varpi \right) \partial _1^{\prime }U_{0,1}\left( \varpi \right) +%
\mathrm{i}\Theta _1^{\prime \prime }+\mathrm{i}\Upsilon _1^{\prime \prime
}\gamma ^{\left[ 5\right] }\right) \\ 
+\beta ^{\left[ 2\right] }\left( \partial _2+U_{0,1}^{-1}\left( \varpi
\right) \partial _2U_{0,1}\left( \varpi \right) +\mathrm{i}\Theta _2+%
\mathrm{i}\Upsilon _2\gamma ^{\left[ 5\right] }\right) \\ 
+\beta ^{\left[ 3\right] }\left( \partial _3+U_{0,1}^{-1}\left( \varpi
\right) \partial _3U_{0,1}\left( \varpi \right) +\mathrm{i}\Theta _3+%
\mathrm{i}\Upsilon _3\gamma ^{\left[ 5\right] }\right) \\ 
+\mathrm{i}M_0\gamma ^{\left[ 0\right] }+\mathrm{i}M_4\beta ^{\left[
4\right] }+\widehat{M}^{\prime \prime }
\end{array}
\right) \varphi =0 
\]

with

\[
\begin{array}{c}
\Theta _0^{\prime \prime }\stackrel{def}{=}\Theta _0\cosh 2\varpi +\Theta
_1\sinh 2\varpi \mbox{,} \\ 
\Theta _1^{\prime \prime }\stackrel{def}{=}\Theta _1\cosh 2\varpi +\Theta
_0\sinh 2\varpi \mbox{,} \\ 
\Upsilon _0^{\prime \prime }\stackrel{def}{=}\Upsilon _0\cosh 2\varpi
+\sinh 2\varpi \cdot \Upsilon _1\mbox{,} \\ 
\Upsilon _1^{\prime \prime }\stackrel{def}{=}\Upsilon _1\cosh 2\varpi
+\Upsilon _0\sinh 2\varpi \mbox{.}
\end{array}
\]

Therefore, the oscillation between blue and green quarks colors with the
oscillation between up and down quarks curves the space of events in the $t$%
, $x_1$ directions.

Similarly that: matrix

\[
U_{0,2}\left( \phi \right) \stackrel{def}{=}\left[ 
\begin{array}{cccc}
\cosh \phi & \mathrm{i}\sinh \phi & 0 & 0 \\ 
-\mathrm{i}\sinh \phi & \cosh \phi & 0 & 0 \\ 
0 & 0 & \cosh \phi & -\mathrm{i}\sinh \phi \\ 
0 & 0 & \mathrm{i}\sinh \phi & \cosh \phi
\end{array}
\right] 
\]

with an arbitrary real function $\phi \left( t,x_1,x_2,x_3\right) $
describes the oscillation between blue and red quarks colors with the
oscillation between up and down quarks which curves the space of events in
the $t$, $x_2$ directions. And matrix

\[
U_{0,3}\left( \iota \right) \stackrel{def}{=}\left[ 
\begin{array}{cccc}
\cosh \iota +\sinh \iota & 0 & 0 & 0 \\ 
0 & \cosh \iota -\sinh \iota & 0 & 0 \\ 
0 & 0 & \cosh \iota -\sinh \iota & 0 \\ 
0 & 0 & 0 & \cosh \iota +\sinh \iota
\end{array}
\right] 
\]

with an arbitrary real function $\iota \left( t,x_1,x_2,x_3\right) $
describes the oscillation between green and red quarks colors with the
oscillation between up and down quarks which curves the space of events in
the $t$, $x_3$ directions.

From (\ref{grg}):

\begin{eqnarray*}
\frac{\partial x_1}{\partial t^{\prime }} &=&\mathrm{c}\sinh 2\varpi \mbox{%
,} \\
\frac{\partial t}{\partial t^{\prime }} &=&\cosh 2\varpi \mbox{.}
\end{eqnarray*}

Because

\begin{eqnarray*}
\sinh 2\varpi  &=&\frac v{\sqrt{1-\frac{v^2}{\mathrm{c}^2}}}\mbox{,} \\
\cosh 2\varpi  &=&\frac 1{\sqrt{1-\frac{v^2}{\mathrm{c}^2}}}
\end{eqnarray*}

with $v$ as an velosity of system $\left\{ t^{\prime },x_1^{\prime }\right\} 
$ as respects to system $\left\{ t,x_1\right\} $ then 

\[
v=\tanh 2\varpi \mbox{.}
\]

Let 
\[
2\varpi :=\nu \left( x_1\right) \frac t{x_1}
\]

with

\[
\nu \left( x_1\right) =\frac \lambda {\left| x_1\right| }\mbox{, }\lambda 
\mbox{ is some
positive real constant.}
\]

In that case

\[
v\left( t,x_1\right) =\tanh \left( \nu \left( x_1\right) \frac t{x_1}\right) 
\]

and if $\mathrm{g}$ is an acseleration of system $\left\{ t^{\prime
},x_1^{\prime }\right\} $ as respects to system $\left\{ t,x_1\right\} $ then

\[
\mathrm{g}\left( t,x_1\right) =\frac{\partial v}{\partial t}=\frac{\nu
\left( x_1\right) }{\left( \cosh ^2\nu \left( x_1\right) \frac t{x_1}\right)
x_1}\mbox{.}
\]

\begin{figure}
\centering
\includegraphics[width=1.0\textwidth]{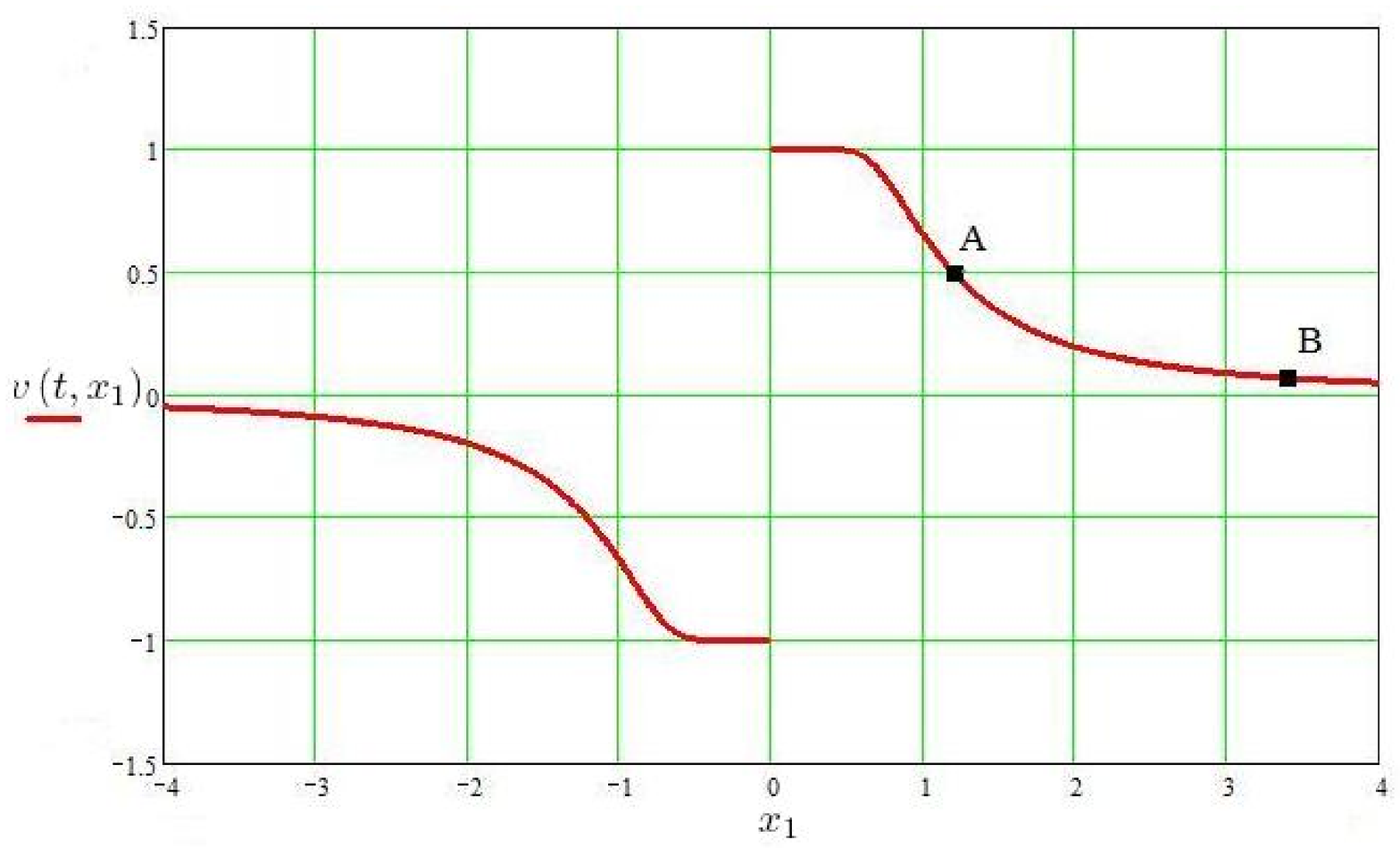}
\caption{}
\end{figure}

\begin{figure}
\centering
\includegraphics[width=1.0\textwidth]{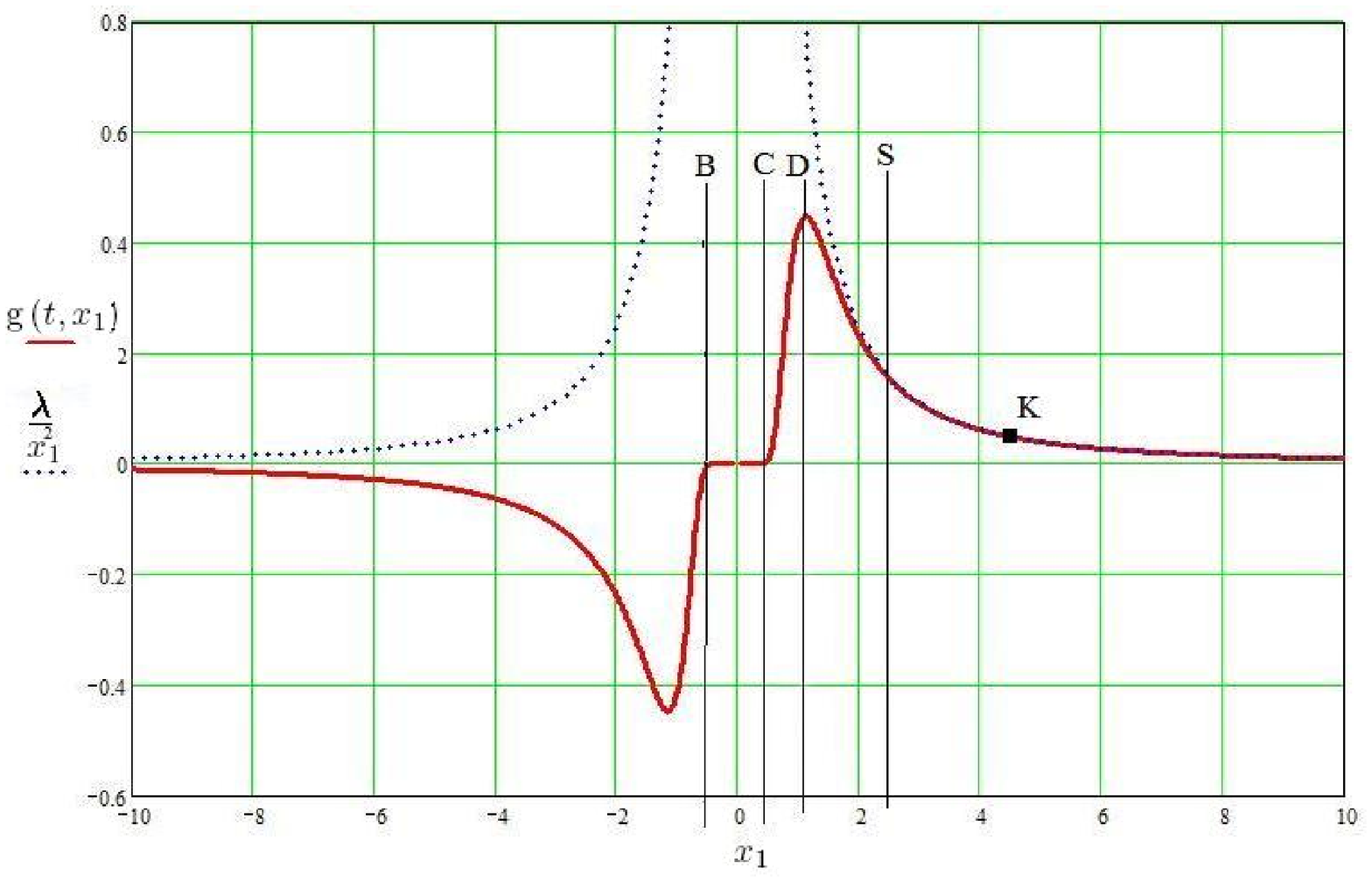}
\caption{}
\end{figure}

Figure 1 shows a dependency from $x_1$ of a system $\left\{ t^{\prime },x_1^{\prime
}\right\} $ velocity $v\left( t,x_1\right) $ in system $\left\{ t,x_1\right\} $.

This velocity in point $A$ does not equal to one in point $B$. Hence, an oscillator, 
placed in $B$, has got a nonzero velosity in respect to an observer, placed in point 
$A$. Therefore, by the Lorentz transformations \cite{Q}, this oscillator frequency for observer, 
placed in point $A$, is less than own frequency of this oscillator ({\it red shift}).

Figure 2 shows a dependency from $x_1$ of a system $\left\{ t^{\prime },x_1^{\prime
}\right\} $ acseleration $g\left( t,x_1\right) $ in system $\left\{ t,x_1\right\} $.

If an immovable in system $\left\{ t,x_1\right\} $ object is placed in point $K$ then 
in system $\left\{ t^{\prime },x_1^{\prime}\right\} $ this object must move to the left  
direction with acseleration $g$ and $\mathrm{g}\simeq \frac \lambda {x_1^2}$. 

Zone from $S$ to $\infty $ is {\it the Newton Gravity Zone}.

Zone from $B$ to $C$ is {\it the Asimptotic Freedom Zone}.

And zone from $C$ to $D$ is {\it the Confinement Force Zone}.

\section{Appendix}

Let

\begin{eqnarray*}
&&\widetilde{U}\left( \chi \right) \stackrel{def}{=} \\
&&\left[ 
\begin{array}{cccc}
\cos \chi +\mathrm{i}\sin \chi & 0 & 0 & 0 \\ 
0 & \cos \chi +\mathrm{i}\sin \chi & 0 & 0 \\ 
0 & 0 & \cos 2\chi +\mathrm{i}\sin 2\chi & 0 \\ 
0 & 0 & 0 & \cos 2\chi +\mathrm{i}\sin 2\chi
\end{array}
\right]
\end{eqnarray*}

and

\[
\widehat{M}^{\prime }\stackrel{def}{=} 
\begin{array}{c}
-M_{\zeta ,0}^{\prime }\gamma _\zeta ^{[0]}+M_{\zeta ,4}^{\prime }\zeta
^{[4]}- \\ 
-M_{\eta ,0}^{\prime }\gamma _\eta ^{[0]}-M_{\eta ,4}^{\prime }\eta ^{[4]}+
\\ 
+M_{\theta ,0}^{\prime }\gamma _\theta ^{[0]}+M_{\theta ,4}^{\prime }\theta
^{[4]}
\end{array}
\stackrel{def}{=}\widetilde{U}^{\dagger }\left( \chi \right) \widehat{M}%
\widetilde{U}\left( \chi \right) 
\]

then:

\begin{eqnarray*}
M_{\zeta ,0}^{\prime } &=&\left( M_{\zeta ,0}\cos \chi -M_{\zeta ,4}\sin
\chi \right) \mbox{,} \\
M_{\zeta ,4}^{\prime } &=&\left( M_{\zeta ,4}\cos \chi +M_{\zeta ,0}\sin
\chi \right) \mbox{,} \\
M_{\eta ,4}^{\prime } &=&\left( M_{\eta ,4}\cos \chi -M_{\eta ,0}\sin \chi
\right) \mbox{,} \\
M_{\eta ,0}^{\prime } &=&\left( M_{\eta ,0}\cos \chi +M_{\eta ,4}\sin \chi
\right) \mbox{,} \\
M_{\theta ,0}^{\prime } &=&\left( M_{\theta ,0}\cos \chi +M_{\theta ,4}\sin
\chi \right) \mbox{,} \\
M_{\theta ,4}^{\prime } &=&\left( M_{\theta ,4}\cos \chi -M_{\theta ,0}\sin
\chi \right) \mbox{.}
\end{eqnarray*}

Therefore, matrix $\widetilde{U}\left( \chi \right) $ makes an oscillation
between up and down quarks.

Let us consider equation (\ref{ham0}) under transformation $\widetilde{U}%
\left( \chi \right) $ with $\chi $ is an arbitrary real function of
time-space variables ($\chi =\chi \left( t,x_1,x_2,x_3\right) $):

\[
\widetilde{U}^{\dagger }\left( \chi \right) \left( \frac 1{\mathrm{c}%
}\partial _t+\mathrm{i}\Theta _0+\mathrm{i}\Upsilon _0\gamma ^{\left[
5\right] }\right) \widetilde{U}\left( \chi \right) \varphi = 
\]

\[
=\widetilde{U}^{\dagger }\left( \chi \right) \left( 
\begin{array}{c}
\beta ^{\left[ 1\right] }\left( \partial _1+\mathrm{i}\Theta _1+\mathrm{i}%
\Upsilon _1\gamma ^{\left[ 5\right] }\right) + \\ 
+\beta ^{\left[ 2\right] }\left( \partial _2+\mathrm{i}\Theta _2+\mathrm{i}%
\Upsilon _2\gamma ^{\left[ 5\right] }\right) + \\ 
+\beta ^{\left[ 3\right] }\left( \partial _3+\mathrm{i}\Theta _3+\mathrm{i}%
\Upsilon _3\gamma ^{\left[ 5\right] }\right) + \\ 
+\widehat{M}
\end{array}
\right) \widetilde{U}\left( \chi \right) \varphi \mbox{.} 
\]

Hence:

\[
\widetilde{U}^{\dagger }\left( \chi \right) \left( \frac 1{\mathrm{c}%
}\partial _t\left( \widetilde{U}\left( \chi \right) \varphi \right) +\mathrm{%
i}\Theta _0\widetilde{U}\left( \chi \right) \varphi +\mathrm{i}\Upsilon
_0\gamma ^{\left[ 5\right] }\widetilde{U}\left( \chi \right) \varphi \right)
= 
\]

\[
=\widetilde{U}^{\dagger }\left( \chi \right) \left( 
\begin{array}{c}
\beta ^{\left[ 1\right] }\left( \partial _1\left( \widetilde{U}\left( \chi
\right) \varphi \right) +\mathrm{i}\Theta _1\widetilde{U}\left( \chi \right)
\varphi +\mathrm{i}\Upsilon _1\gamma ^{\left[ 5\right] }\widetilde{U}\left(
\chi \right) \varphi \right) + \\ 
+\beta ^{\left[ 2\right] }\left( \partial _2\left( \widetilde{U}\left( \chi
\right) \varphi \right) +\mathrm{i}\Theta _2\widetilde{U}\left( \chi \right)
\varphi +\mathrm{i}\Upsilon _2\gamma ^{\left[ 5\right] }\widetilde{U}\left(
\chi \right) \varphi \right) + \\ 
+\beta ^{\left[ 3\right] }\left( \partial _3\left( \widetilde{U}\left( \chi
\right) \varphi \right) +\mathrm{i}\Theta _3\widetilde{U}\left( \chi \right)
\varphi +\mathrm{i}\Upsilon _3\gamma ^{\left[ 5\right] }\widetilde{U}\left(
\chi \right) \varphi \right) + \\ 
+\widehat{M}\widetilde{U}\left( \chi \right) \varphi
\end{array}
\right) \mbox{.} 
\]

Because

\[
\gamma ^{\left[ 5\right] }\widetilde{U}\left( \chi \right) =\widetilde{U}%
\left( \chi \right) \gamma ^{\left[ 5\right] } 
\]

then

\[
\widetilde{U}^{\dagger }\left( \chi \right) \left( 
\begin{array}{c}
\frac 1{\mathrm{c}}\left( \partial _t\widetilde{U}\left( \chi \right)
\right) \varphi +\widetilde{U}\left( \chi \right) \frac 1{\mathrm{c}%
}\partial _t\varphi \\ 
+\mathrm{i}\Theta _0\widetilde{U}\left( \chi \right) \varphi +\widetilde{U}%
\left( \chi \right) \mathrm{i}\Upsilon _0\gamma ^{\left[ 5\right] }\varphi
\end{array}
\right) = 
\]

\[
=\widetilde{U}^{\dagger }\left( \chi \right) \left( 
\begin{array}{c}
\beta ^{\left[ 1\right] }\left( 
\begin{array}{c}
\left( \partial _1\widetilde{U}\left( \chi \right) \right) \varphi +%
\widetilde{U}\left( \chi \right) \partial _1\varphi \\ 
+\mathrm{i}\Theta _1\widetilde{U}\left( \chi \right) \varphi +\widetilde{U}%
\left( \chi \right) \mathrm{i}\Upsilon _1\gamma ^{\left[ 5\right] }\varphi
\end{array}
\right) + \\ 
+\beta ^{\left[ 2\right] }\left( 
\begin{array}{c}
\left( \partial _2\widetilde{U}\left( \chi \right) \right) \varphi +%
\widetilde{U}\left( \chi \right) \partial _2\varphi \\ 
+\mathrm{i}\Theta _2\widetilde{U}\left( \chi \right) \varphi +\widetilde{U}%
\left( \chi \right) \mathrm{i}\Upsilon _2\gamma ^{\left[ 5\right] }\varphi
\end{array}
\right) + \\ 
+\beta ^{\left[ 3\right] }\left( 
\begin{array}{c}
\left( \partial _3\widetilde{U}\left( \chi \right) \right) \varphi +%
\widetilde{U}\left( \chi \right) \partial _3\varphi \\ 
+\mathrm{i}\Theta _3\widetilde{U}\left( \chi \right) \varphi +\widetilde{U}%
\left( \chi \right) \mathrm{i}\Upsilon _3\gamma ^{\left[ 5\right] }\varphi
\end{array}
\right) + \\ 
+\widehat{M}\widetilde{U}\left( \chi \right) \varphi
\end{array}
\right) \mbox{.} 
\]

Because

\begin{eqnarray*}
\beta ^{\left[ 1\right] }\widetilde{U}\left( \chi \right) &=&\widetilde{U}%
\left( \chi \right) \beta ^{\left[ 1\right] }\mbox{,} \\
\beta ^{\left[ 2\right] }\widetilde{U}\left( \chi \right) &=&\widetilde{U}%
\left( \chi \right) \beta ^{\left[ 2\right] }\mbox{,} \\
\beta ^{\left[ 3\right] }\widetilde{U}\left( \chi \right) &=&\widetilde{U}%
\left( \chi \right) \beta ^{\left[ 3\right] }
\end{eqnarray*}

then:

\[
\left( 
\begin{array}{c}
+\widetilde{U}^{\dagger }\left( \chi \right) \widetilde{U}\left( \chi
\right) \frac 1{\mathrm{c}}\partial _t+\frac 1{\mathrm{c}}\widetilde{U}%
^{\dagger }\left( \chi \right) \left( \partial _t\widetilde{U}\left( \chi
\right) \right) \\ 
+\mathrm{i}\Theta _0\widetilde{U}^{\dagger }\left( \chi \right) \widetilde{U}%
\left( \chi \right) +\widetilde{U}^{\dagger }\left( \chi \right) \widetilde{U%
}\left( \chi \right) \mathrm{i}\Upsilon _0\gamma ^{\left[ 5\right] }
\end{array}
\right) \varphi = 
\]

\[
=\left( 
\begin{array}{c}
\beta ^{\left[ 1\right] }\left( 
\begin{array}{c}
\widetilde{U}^{\dagger }\left( \chi \right) \widetilde{U}\left( \chi \right)
\partial _1+\widetilde{U}^{\dagger }\left( \chi \right) \left( \partial _1%
\widetilde{U}\left( \chi \right) \right) \\ 
+\mathrm{i}\Theta _1\widetilde{U}^{\dagger }\left( \chi \right) \widetilde{U}%
\left( \chi \right) +\widetilde{U}^{\dagger }\left( \chi \right) \widetilde{U%
}\left( \chi \right) \mathrm{i}\Upsilon _1\gamma ^{\left[ 5\right] }
\end{array}
\right) + \\ 
+\beta ^{\left[ 2\right] }\left( 
\begin{array}{c}
\widetilde{U}^{\dagger }\left( \chi \right) \widetilde{U}\left( \chi \right)
\partial _2+\widetilde{U}^{\dagger }\left( \chi \right) \left( \partial _2%
\widetilde{U}\left( \chi \right) \right) \\ 
+\mathrm{i}\Theta _2\widetilde{U}^{\dagger }\left( \chi \right) \widetilde{U}%
\left( \chi \right) +\widetilde{U}^{\dagger }\left( \chi \right) \widetilde{U%
}\left( \chi \right) \mathrm{i}\Upsilon _2\gamma ^{\left[ 5\right] }
\end{array}
\right) + \\ 
+\beta ^{\left[ 3\right] }\left( 
\begin{array}{c}
\widetilde{U}^{\dagger }\left( \chi \right) \widetilde{U}\left( \chi \right)
\partial _3+\widetilde{U}^{\dagger }\left( \chi \right) \left( \partial _3%
\widetilde{U}\left( \chi \right) \right) \\ 
+\mathrm{i}\Theta _3\widetilde{U}^{\dagger }\left( \chi \right) \widetilde{U}%
\left( \chi \right) +\widetilde{U}^{\dagger }\left( \chi \right) \widetilde{U%
}\left( \chi \right) \mathrm{i}\Upsilon _3\gamma ^{\left[ 5\right] }
\end{array}
\right) + \\ 
+\widetilde{U}^{\dagger }\left( \chi \right) \widehat{M}\widetilde{U}\left(
\chi \right)
\end{array}
\right) \varphi \mbox{.} 
\]

Because:

\[
\widetilde{U}^{\dagger }\left( \chi \right) \widetilde{U}\left( \chi \right)
=1_4 
\]

then

\[
\left( \frac 1{\mathrm{c}}\partial _t+\frac 1{\mathrm{c}}\widetilde{U}%
^{\dagger }\left( \chi \right) \left( \partial _t\widetilde{U}\left( \chi
\right) \right) +\mathrm{i}\Theta _0+\mathrm{i}\Upsilon _0\gamma ^{\left[
5\right] }\right) \varphi = 
\]

\[
=\left( 
\begin{array}{c}
\beta ^{\left[ 1\right] }\left( \partial _1+\widetilde{U}^{\dagger }\left(
\chi \right) \left( \partial _1\widetilde{U}\left( \chi \right) \right) +%
\mathrm{i}\Theta _1+\mathrm{i}\Upsilon _1\gamma ^{\left[ 5\right] }\right) +
\\ 
+\beta ^{\left[ 2\right] }\left( \partial _2+\widetilde{U}^{\dagger }\left(
\chi \right) \left( \partial _2\widetilde{U}\left( \chi \right) \right) +%
\mathrm{i}\Theta _2+\mathrm{i}\Upsilon _2\gamma ^{\left[ 5\right] }\right) +
\\ 
+\beta ^{\left[ 3\right] }\left( \partial _3+\widetilde{U}^{\dagger }\left(
\chi \right) \left( \partial _3\widetilde{U}\left( \chi \right) \right) +%
\mathrm{i}\Theta _3+\mathrm{i}\Upsilon _3\gamma ^{\left[ 5\right] }\right) +
\\ 
+\widetilde{U}^{\dagger }\left( \chi \right) \widehat{M}\widetilde{U}\left(
\chi \right)
\end{array}
\right) \varphi \mbox{.} 
\]

Now let:

\[
\begin{array}{c}
\widehat{U}\left( \kappa \right) \stackrel{def}{=} \\ 
\left[ 
\begin{array}{cccc}
\cosh \kappa +\sinh \kappa & 0 & 0 & 0 \\ 
0 & \cosh \kappa +\sinh \kappa & 0 & 0 \\ 
0 & 0 & \cosh 2\kappa +\sinh 2\kappa & 0 \\ 
0 & 0 & 0 & \cosh 2\kappa +\sinh 2\kappa
\end{array}
\right]
\end{array}
\]

and

\[
\widehat{M}^{\prime }\stackrel{def}{=} 
\begin{array}{c}
-M_{\zeta ,0}^{\prime }\gamma _\zeta ^{[0]}+M_{\zeta ,4}^{\prime }\zeta
^{[4]}- \\ 
-M_{\eta ,0}^{\prime }\gamma _\eta ^{[0]}-M_{\eta ,4}^{\prime }\eta ^{[4]}+
\\ 
+M_{\theta ,0}^{\prime }\gamma _\theta ^{[0]}+M_{\theta ,4}^{\prime }\theta
^{[4]}
\end{array}
\stackrel{def}{=}\widehat{U}^{-1}\left( \kappa \right) \widehat{M}\widehat{U}%
\left( \kappa \right) 
\]

then:

\begin{eqnarray*}
M_{\theta ,0}^{\prime } &=&\left( M_{\theta ,0}\cosh \kappa -\mathrm{i}%
M_{\theta ,4}\sinh \kappa \right) \mbox{,} \\
M_{\theta ,4}^{\prime } &=&\left( M_{\theta ,4}\cosh \kappa +\mathrm{i}%
M_{\theta ,0}\sinh \kappa \right) \mbox{,} \\
M_{\eta ,0}^{\prime } &=&\left( M_{\eta ,0}\cosh \kappa -\mathrm{i}M_{\eta
,4}\sinh \kappa \right) \mbox{,} \\
M_{\eta ,4}^{\prime } &=&\left( M_{\eta ,4}\cosh \kappa +\mathrm{i}M_{\eta
,0}\sinh \kappa \right) \mbox{,} \\
M_{\zeta ,0}^{\prime } &=&\left( M_{\zeta ,0}\cosh \kappa +\mathrm{i}%
M_{\zeta ,4}\sinh \kappa \right) \mbox{,} \\
M_{\zeta ,4}^{\prime } &=&\left( M_{\zeta ,4}\cosh \kappa -\mathrm{i}%
M_{\zeta ,0}\sinh \kappa \right) \mbox{.}
\end{eqnarray*}

Therefore, matrix $\widehat{U}\left( \kappa \right) $ makes an oscillation
between up and down quarks, too.

Let us consider equation (\ref{ham0}) under transformation $\widehat{U}%
\left( \kappa \right) $ with $\kappa $ is an arbitrary real function of
time-space variables ($\kappa =\kappa \left( t,x_1,x_2,x_3\right) $):

\[
\widehat{U}^{-1}\left( \kappa \right) \left( \frac 1{\mathrm{c}}\partial _t+%
\mathrm{i}\Theta _0+\mathrm{i}\Upsilon _0\gamma ^{\left[ 5\right] }\right) 
\widehat{U}\left( \kappa \right) \varphi = 
\]

\[
=\widehat{U}^{-1}\left( \kappa \right) \left( 
\begin{array}{c}
\beta ^{\left[ 1\right] }\left( \partial _1+\mathrm{i}\Theta _1+\mathrm{i}%
\Upsilon _1\gamma ^{\left[ 5\right] }\right) + \\ 
+\beta ^{\left[ 2\right] }\left( \partial _2+\mathrm{i}\Theta _2+\mathrm{i}%
\Upsilon _2\gamma ^{\left[ 5\right] }\right) + \\ 
+\beta ^{\left[ 3\right] }\left( \partial _3+\mathrm{i}\Theta _3+\mathrm{i}%
\Upsilon _3\gamma ^{\left[ 5\right] }\right) + \\ 
+\widehat{M}
\end{array}
\right) \widehat{U}\left( \kappa \right) \varphi \mbox{.} 
\]

Hence:

\[
\widehat{U}^{-1}\left( \kappa \right) \left( \frac 1{\mathrm{c}}\partial
_t\left( \widehat{U}\left( \kappa \right) \varphi \right) +\mathrm{i}\Theta
_0\widehat{U}\left( \kappa \right) \varphi +\mathrm{i}\Upsilon _0\gamma
^{\left[ 5\right] }\widehat{U}\left( \kappa \right) \varphi \right) = 
\]

\[
=\widehat{U}^{-1}\left( \kappa \right) \left( 
\begin{array}{c}
\beta ^{\left[ 1\right] }\left( \partial _1\left( \widehat{U}\left( \kappa
\right) \varphi \right) +\mathrm{i}\Theta _1\widehat{U}\left( \kappa \right)
\varphi +\mathrm{i}\Upsilon _1\gamma ^{\left[ 5\right] }\widehat{U}\left(
\kappa \right) \varphi \right) + \\ 
+\beta ^{\left[ 2\right] }\left( \partial _2\left( \widehat{U}\left( \kappa
\right) \varphi \right) +\mathrm{i}\Theta _2\widehat{U}\left( \kappa \right)
\varphi +\mathrm{i}\Upsilon _2\gamma ^{\left[ 5\right] }\widehat{U}\left(
\kappa \right) \varphi \right) + \\ 
+\beta ^{\left[ 3\right] }\left( \partial _3\left( \widehat{U}\left( \kappa
\right) \varphi \right) +\mathrm{i}\Theta _3\widehat{U}\left( \kappa \right)
\varphi +\mathrm{i}\Upsilon _3\gamma ^{\left[ 5\right] }\widehat{U}\left(
\kappa \right) \varphi \right) + \\ 
+\widehat{M}\widehat{U}\left( \kappa \right) \varphi
\end{array}
\right) \mbox{.} 
\]

Because

\[
\gamma ^{\left[ 5\right] }\widehat{U}\left( \kappa \right) =\widehat{U}%
\left( \kappa \right) \gamma ^{\left[ 5\right] } 
\]

then

\[
\widehat{U}^{-1}\left( \kappa \right) \left( 
\begin{array}{c}
\left( \frac 1{\mathrm{c}}\partial _t\widehat{U}\left( \kappa \right)
\right) \varphi +\widehat{U}\left( \kappa \right) \frac 1{\mathrm{c}%
}\partial _t\varphi \\ 
+\mathrm{i}\Theta _0\widehat{U}\left( \kappa \right) \varphi +\mathrm{i}%
\Upsilon _0\widehat{U}\left( \kappa \right) \gamma ^{\left[ 5\right] }\varphi
\end{array}
\right) = 
\]

\[
=\widehat{U}^{-1}\left( \kappa \right) \left( 
\begin{array}{c}
\beta ^{\left[ 1\right] }\left( 
\begin{array}{c}
\left( \partial _1\widehat{U}\left( \kappa \right) \right) \varphi +\widehat{%
U}\left( \kappa \right) \partial _1\varphi \\ 
+\mathrm{i}\Theta _1\widehat{U}\left( \kappa \right) \varphi +\mathrm{i}%
\Upsilon _1\widehat{U}\left( \kappa \right) \gamma ^{\left[ 5\right] }\varphi
\end{array}
\right) + \\ 
+\beta ^{\left[ 2\right] }\left( 
\begin{array}{c}
\left( \partial _2\widehat{U}\left( \kappa \right) \right) \varphi +\widehat{%
U}\left( \kappa \right) \partial _2\varphi \\ 
+\mathrm{i}\Theta _2\widehat{U}\left( \kappa \right) \varphi +\mathrm{i}%
\Upsilon _2\widehat{U}\left( \kappa \right) \gamma ^{\left[ 5\right] }\varphi
\end{array}
\right) + \\ 
+\beta ^{\left[ 3\right] }\left( 
\begin{array}{c}
\left( \partial _3\widehat{U}\left( \kappa \right) \right) \varphi +\widehat{%
U}\left( \kappa \right) \partial _3\varphi \\ 
+\mathrm{i}\Theta _3\widehat{U}\left( \kappa \right) \varphi +\mathrm{i}%
\Upsilon _3\widehat{U}\left( \kappa \right) \gamma ^{\left[ 5\right] }\varphi
\end{array}
\right) + \\ 
+\widehat{M}\widehat{U}\left( \kappa \right) \varphi
\end{array}
\right) \mbox{.} 
\]

Because

\begin{eqnarray*}
\widehat{U}^{-1}\left( \kappa \right) \beta ^{\left[ 1\right] } &=&\beta
^{\left[ 1\right] }\widehat{U}^{-1}\left( \kappa \right) \mbox{,} \\
\widehat{U}^{-1}\left( \kappa \right) \beta ^{\left[ 2\right] } &=&\beta
^{\left[ 2\right] }\widehat{U}^{-1}\left( \kappa \right) \mbox{,} \\
\widehat{U}^{-1}\left( \kappa \right) \beta ^{\left[ 3\right] } &=&\beta
^{\left[ 3\right] }\widehat{U}^{-1}\left( \kappa \right)
\end{eqnarray*}

then

\[
\left( 
\begin{array}{c}
\widehat{U}^{-1}\left( \kappa \right) \left( \frac 1{\mathrm{c}}\partial _t%
\widehat{U}\left( \kappa \right) \right) +\widehat{U}^{-1}\left( \kappa
\right) \widehat{U}\left( \kappa \right) \frac 1{\mathrm{c}}\partial _t \\ 
+\mathrm{i}\Theta _0\widehat{U}^{-1}\left( \kappa \right) \widehat{U}\left(
\kappa \right) +\mathrm{i}\Upsilon _0\widehat{U}^{-1}\left( \kappa \right) 
\widehat{U}\left( \kappa \right) \gamma ^{\left[ 5\right] }
\end{array}
\right) \varphi = 
\]

\[
=\left( 
\begin{array}{c}
\beta ^{\left[ 1\right] }\left( 
\begin{array}{c}
\widehat{U}^{-1}\left( \kappa \right) \left( \partial _1\widehat{U}\left(
\kappa \right) \right) +\widehat{U}^{-1}\left( \kappa \right) \widehat{U}%
\left( \kappa \right) \partial _1 \\ 
+\mathrm{i}\Theta _1\widehat{U}^{-1}\left( \kappa \right) \widehat{U}\left(
\kappa \right) +\mathrm{i}\Upsilon _1\widehat{U}^{-1}\left( \kappa \right) 
\widehat{U}\left( \kappa \right) \gamma ^{\left[ 5\right] }
\end{array}
\right) + \\ 
+\beta ^{\left[ 2\right] }\left( 
\begin{array}{c}
\widehat{U}^{-1}\left( \kappa \right) \left( \partial _2\widehat{U}\left(
\kappa \right) \right) +\widehat{U}^{-1}\left( \kappa \right) \widehat{U}%
\left( \kappa \right) \partial _2 \\ 
+\mathrm{i}\Theta _2\widehat{U}^{-1}\left( \kappa \right) \widehat{U}\left(
\kappa \right) +\mathrm{i}\Upsilon _2\widehat{U}^{-1}\left( \kappa \right) 
\widehat{U}\left( \kappa \right) \gamma ^{\left[ 5\right] }
\end{array}
\right) + \\ 
+\beta ^{\left[ 3\right] }\left( 
\begin{array}{c}
\widehat{U}^{-1}\left( \kappa \right) \left( \partial _3\widehat{U}\left(
\kappa \right) \right) +\widehat{U}^{-1}\left( \kappa \right) \widehat{U}%
\left( \kappa \right) \partial _3 \\ 
+\mathrm{i}\Theta _3\widehat{U}^{-1}\left( \kappa \right) \widehat{U}\left(
\kappa \right) +\mathrm{i}\Upsilon _3\widehat{U}^{-1}\left( \kappa \right) 
\widehat{U}\left( \kappa \right) \gamma ^{\left[ 5\right] }
\end{array}
\right) + \\ 
+\widehat{U}^{-1}\left( \kappa \right) \widehat{M}\widehat{U}\left( \kappa
\right)
\end{array}
\right) \varphi \mbox{.} 
\]

Because

\[
\widehat{U}^{-1}\left( \kappa \right) \widehat{U}\left( \kappa \right) =1_4 
\]

then

\[
\left( \frac 1{\mathrm{c}}\partial _t+\widehat{U}^{-1}\left( \kappa \right)
\left( \frac 1{\mathrm{c}}\partial _t\widehat{U}\left( \kappa \right)
\right) +\mathrm{i}\Theta _0+\mathrm{i}\Upsilon _0\gamma ^{\left[ 5\right]
}\right) \varphi = 
\]

\[
=\left( 
\begin{array}{c}
\beta ^{\left[ 1\right] }\left( \partial _1+\widehat{U}^{-1}\left( \kappa
\right) \left( \partial _1\widehat{U}\left( \kappa \right) \right) +\mathrm{i%
}\Theta _1+\mathrm{i}\Upsilon _1\gamma ^{\left[ 5\right] }\right) + \\ 
+\beta ^{\left[ 2\right] }\left( \partial _2+\widehat{U}^{-1}\left( \kappa
\right) \left( \partial _2\widehat{U}\left( \kappa \right) \right) +\mathrm{i%
}\Theta _2+\mathrm{i}\Upsilon _2\gamma ^{\left[ 5\right] }\right) + \\ 
+\beta ^{\left[ 3\right] }\left( \partial _3+\widehat{U}^{-1}\left( \kappa
\right) \left( \partial _3\widehat{U}\left( \kappa \right) \right) +\mathrm{i%
}\Theta _3+\mathrm{i}\Upsilon _3\gamma ^{\left[ 5\right] }\right) + \\ 
+\widehat{U}^{-1}\left( \kappa \right) \widehat{M}\widehat{U}\left( \kappa
\right)
\end{array}
\right) \varphi \mbox{.} 
\]

If denote:

\begin{eqnarray*}
&&\ \ \Lambda _1\stackrel{def}{=}\left[ 
\begin{array}{cccc}
0 & -1 & 0 & 0 \\ 
-1 & 0 & 0 & 0 \\ 
0 & 0 & 0 & 1 \\ 
0 & 0 & 1 & 0
\end{array}
\right] \mbox{, }\Lambda _2\stackrel{def}{=}\left[ 
\begin{array}{cccc}
0 & \mathrm{i} & 0 & 0 \\ 
\mathrm{i} & 0 & 0 & 0 \\ 
0 & 0 & 0 & \mathrm{i} \\ 
0 & 0 & \mathrm{i} & 0
\end{array}
\right] \mbox{, } \\
&&\ \ \Lambda _3\stackrel{def}{=}\left[ 
\begin{array}{cccc}
0 & 1 & 0 & 0 \\ 
-1 & 0 & 0 & 0 \\ 
0 & 0 & 0 & 1 \\ 
0 & 0 & -1 & 0
\end{array}
\right] \mbox{, }\Lambda _4\stackrel{def}{=}\left[ 
\begin{array}{cccc}
0 & \mathrm{i} & 0 & 0 \\ 
-\mathrm{i} & 0 & 0 & 0 \\ 
0 & 0 & 0 & -\mathrm{i} \\ 
0 & 0 & \mathrm{i} & 0
\end{array}
\right] \mbox{,} \\
&&\ \Lambda _5\stackrel{def}{=}\left[ 
\begin{array}{cccc}
-\mathrm{i} & 0 & 0 & 0 \\ 
0 & \mathrm{i} & 0 & 0 \\ 
0 & 0 & -\mathrm{i} & 0 \\ 
0 & 0 & 0 & \mathrm{i}
\end{array}
\right] \mbox{, }\Lambda _6\stackrel{def}{=}\left[ 
\begin{array}{cccc}
1 & 0 & 0 & 0 \\ 
0 & -1 & 0 & 0 \\ 
0 & 0 & -1 & 0 \\ 
0 & 0 & 0 & 1
\end{array}
\right] \mbox{,}
\end{eqnarray*}

\[
\Lambda _7\stackrel{def}{=}\left[ 
\begin{array}{cccc}
1 & 0 & 0 & 0 \\ 
0 & 1 & 0 & 0 \\ 
0 & 0 & 2 & 0 \\ 
0 & 0 & 0 & 2
\end{array}
\right] \mbox{, }\Lambda _8\stackrel{def}{=}\left[ 
\begin{array}{cccc}
\mathrm{i} & 0 & 0 & 0 \\ 
0 & \mathrm{i} & 0 & 0 \\ 
0 & 0 & 2\mathrm{i} & 0 \\ 
0 & 0 & 0 & 2\mathrm{i}
\end{array}
\right] 
\]

then

\begin{eqnarray*}
U_{0,1}^{-1}\left( \varpi \right) \left( \partial _sU_{0,1}\left( \varpi
\right) \right) &=&\Lambda _1\partial _s\varpi \mbox{,} \\
U_{2,3}^{-1}\left( \alpha \right) \left( \partial _sU_{2,3}\left( \alpha
\right) \right) &=&\Lambda _2\partial _s\alpha \mbox{,} \\
U_{1,3}^{-1}\left( \vartheta \right) \left( \partial _sU_{1,3}\left(
\vartheta \right) \right) &=&\Lambda _3\partial _s\vartheta \mbox{,} \\
U_{0,2}^{-1}\left( \phi \right) \left( \partial _sU_{0,2}\left( \phi \right)
\right) &=&\Lambda _4\partial _s\phi \mbox{,} \\
U_{1,2}^{-1}\left( \varsigma \right) \left( \partial _sU_{1,2}\left(
\varsigma \right) \right) &=&\Lambda _5\partial _s\varsigma \mbox{,} \\
U_{0,3}^{-1}\left( \iota \right) \left( \partial _sU_{0,3}\left( \iota
\right) \right) &=&\Lambda _6\partial _s\iota \mbox{,} \\
\widehat{U}^{-1}\left( \kappa \right) \left( \partial _s\widehat{U}\left(
\kappa \right) \right) &=&\Lambda _7\partial _s\kappa \mbox{,} \\
\widetilde{U}^{-1}\left( \chi \right) \left( \partial _s\widetilde{U}\left(
\chi \right) \right) &=&\Lambda _8\partial _s\chi \mbox{.}
\end{eqnarray*}

Let $U$ is a product of elements of the following set:

\[
\grave U\stackrel{def}{=}\left\{
U_{0,1},U_{2,3},U_{1,3},U_{0,2},U_{1,2},U_{0,3},\widehat{U},\widetilde{U}%
\right\} \mbox{.} 
\]

For example, let $U\left( \varsigma ,\kappa ,\iota \right) \stackrel{def}{=}%
U_{0,1}\left( \varpi \right) U_{0,2}\left( \phi \right) U_{1,3}\left(
\vartheta \right) $

In that case

\[
\begin{array}{c}
U^{-1}\left( \varsigma ,\kappa ,\iota \right) \left( \partial _sU\left(
\varsigma ,\kappa ,\iota \right) \right) = \\ 
\left( U_{0,1}\left( \varpi \right) U_{0,2}\left( \phi \right)
U_{1,3}\left( \vartheta \right) \right) ^{-1}\left( \partial _sU_{0,1}\left(
\varpi \right) U_{0,2}\left( \phi \right) U_{1,3}\left( \vartheta \right)
\right) \mbox{.}
\end{array}
\]

Hence:

\[
\begin{array}{c}
U^{-1}\left( \varsigma ,\kappa ,\iota \right) \left( \partial _sU\left(
\varsigma ,\kappa ,\iota \right) \right) = \\ 
U_{1,3}^{-1}\left( \vartheta \right) U_{0,2}^{-1}\left( \phi \right) \Lambda
_1\partial _s\varpi U_{0,2}\left( \phi \right) U_{1,3}\left( \vartheta
\right) \\ 
+U_{1,3}^{-1}\left( \vartheta \right) U_{0,2}^{-1}\left( \phi \right) \left(
\partial _sU_{0,2}\left( \phi \right) \right) U_{1,3}\left( \vartheta \right)
\\ 
+U_{1,3}^{-1}\left( \vartheta \right) \partial _sU_{1,3}\left( \vartheta
\right) \mbox{.}
\end{array}
\]

Therefore:

\[
\begin{array}{c}
U^{-1}\left( \varsigma ,\kappa ,\iota \right) \left( \partial _sU\left(
\varsigma ,\kappa ,\iota \right) \right) = \\ 
U_{1,3}^{-1}\left( \vartheta \right) \left( \Lambda _1\cosh 2\phi -\Lambda
_5\sinh 2\phi \right) U_{1,3}\left( \vartheta \right) \partial _s\varpi \\ 
+U_{1,3}^{-1}\left( \vartheta \right) \Lambda _4\partial _s\phi
U_{1,3}\left( \vartheta \right) \\ 
+\Lambda _3\partial _s\vartheta \mbox{.}
\end{array}
\]

Hence:

\[
\begin{array}{c}
U^{-1}\left( \varsigma ,\kappa ,\iota \right) \left( \partial _sU\left(
\varsigma ,\kappa ,\iota \right) \right) = \\ 
U_{1,3}^{-1}\left( \vartheta \right) \Lambda _1\cosh 2\phi U_{1,3}\left(
\vartheta \right) \partial _s\varpi \\ 
-U_{1,3}^{-1}\left( \vartheta \right) \Lambda _5\sinh 2\phi U_{1,3}\left(
\vartheta \right) \partial _s\varpi \\ 
+\Lambda _4\partial _s\phi +\Lambda _3\partial _s\vartheta \mbox{.}
\end{array}
\]

Hence:

\[
\begin{array}{c}
U^{-1}\left( \varsigma ,\kappa ,\iota \right) \left( \partial _sU\left(
\varsigma ,\kappa ,\iota \right) \right) = \\ 
\left( \Lambda _1\cos 2\vartheta +\Lambda _6\sin 2\vartheta \right) \cosh
2\phi \partial _s\varpi \\ 
-\left( \Lambda _5\cos 2\vartheta -\Lambda _2\sin 2\vartheta \right) \sinh
2\phi \partial _s\varpi \\ 
+\Lambda _4\partial _s\phi +\Lambda _3\partial _s\vartheta \mbox{.}
\end{array}
\]

Hence:

\[
\begin{array}{c}
U^{-1}\left( \varsigma ,\kappa ,\iota \right) \left( \partial _sU\left(
\varsigma ,\kappa ,\iota \right) \right) = \\ 
\Lambda _1\cos 2\vartheta \cosh 2\phi \partial _s\varpi +\Lambda _6\sin
2\vartheta \cosh 2\phi \partial _s\varpi \\ 
-\Lambda _5\cos 2\vartheta \sinh 2\phi \partial _s\varpi -\Lambda _2\sin
2\vartheta \sinh 2\phi \partial _s\varpi \\ 
+\Lambda _4\partial _s\phi +\Lambda _3\partial _s\vartheta \mbox{.}
\end{array}
\]

\[
\begin{array}{c}
U^{-1}\left( \varsigma ,\kappa ,\iota \right) \left( \partial _sU\left(
\varsigma ,\kappa ,\iota \right) \right) = \\ 
\Lambda _1\cos 2\vartheta \cosh 2\phi \partial _s\varpi -\Lambda _2\sin
2\vartheta \sinh 2\phi \partial _s\varpi +\Lambda _3\partial _s\vartheta
\\ 
+\Lambda _4\partial _s\phi -\Lambda _5\cos 2\vartheta \sinh 2\phi \partial
_s\varpi +\Lambda _6\sin 2\vartheta \cosh 2\phi \partial _s\varpi %
\mbox{.}
\end{array}
\]

Similarly that, there for every product $U$ of $\grave U$'s elements real
functions $G_s^r\left( t,x_1,x_2,x_3\right) $ exist such that

\[
U^{-1}\left( \partial _sU\right) =\frac{g_3}2\sum_{r=1}^8\Lambda _rG_s^r
\]

with some real constant $g_3$ (similar to 8 gluons).

---------------------------------------------------------------------------------

$U_{2,3}^{-1}\left( \alpha \right) \Lambda _1U_{2,3}\left( \alpha \right)
=\Lambda _1$

$U_{1,3}^{-1}\left( \vartheta \right) \Lambda _1U_{1,3}\left( \vartheta
\right) =\left( \Lambda _1\cos 2\vartheta +\Lambda _6\sin 2\vartheta \right) 
$

$U_{0,2}^{-1}\left( \phi \right) \Lambda _1U_{0,2}\left( \phi \right)
=\left( \Lambda _1\cosh 2\phi -\Lambda _5\sinh 2\phi \right) $

$U_{1,2}^{-1}\left( \varsigma \right) \Lambda _1U_{1,2}\left( \varsigma
\right) =\Lambda _1\cos 2\varsigma -\Lambda _4\sin 2\varsigma $

$U_{0,3}^{-1}\left( \iota \right) \Lambda _1U_{0,3}\left( \iota \right)
=\Lambda _1\cosh 2\iota +\Lambda _3\sinh 2\iota $

$\widehat{U}^{-1}\left( \kappa \right) \Lambda _1\widehat{U}\left( \kappa
\right) =\Lambda _1$

$\widetilde{U}^{-1}\left( \chi \right) \Lambda _1\widetilde{U}\left( \chi
\right) =\Lambda _1$

========

$\widetilde{U}^{-1}\left( \chi \right) \Lambda _2\widetilde{U}\left( \chi
\right) =\Lambda _2$

$\widehat{U}^{-1}\left( \kappa \right) \Lambda _2\widehat{U}\left( \kappa
\right) =\Lambda _2$

$U_{0,3}^{-1}\left( \iota \right) \Lambda _2U_{0,3}\left( \iota \right)
=\Lambda _2\cosh 2\iota -\Lambda _4\sinh 2\iota $

$U_{1,2}^{-1}\left( \varsigma \right) \Lambda _2U_{1,2}\left( \varsigma
\right) =\Lambda _2\cos 2\varsigma -\Lambda _3\sin 2\varsigma $

$U_{0,2}^{-1}\left( \phi \right) \Lambda _2U_{0,2}\left( \phi \right)
=\Lambda _2\cosh 2\phi +\Lambda _6\sinh 2\phi $

$U_{1,3}^{-1}\left( \vartheta \right) \Lambda _2U_{1,3}\left( \vartheta
\right) =\Lambda _2\cos 2\vartheta +\Lambda _5\sin 2\vartheta $

$U_{0,1}^{-1}\left( \varpi \right) \Lambda _2U_{0,1}\left( \varpi
\right) =\Lambda _2$

=========

$U_{0,1}^{-1}\left( \varpi \right) \ \Lambda _3U_{0,1}\left( \varpi
\right) =\Lambda _3\cosh 2\varpi -\Lambda _6\sinh 2\varpi $

$U_{2,3}^{-1}\left( \alpha \right) \ \Lambda _3U_{2,3}\left( \alpha \right)
=\Lambda _3\cos 2\alpha -\Lambda _5\sin 2\alpha $

$U_{0,2}^{-1}\left( \phi \right) \ \Lambda _3U_{0,2}\left( \phi \right)
=\Lambda _3$

$U_{1,2}^{-1}\left( \varsigma \right) \ \Lambda _3U_{1,2}\left( \varsigma
\right) =\Lambda _3\cos 2\varsigma +\Lambda _2\sin 2\varsigma $

$U_{0,3}^{-1}\left( \iota \right) \ \Lambda _3U_{0,3}\left( \iota \right) =\
\Lambda _3\cosh 2\iota +\Lambda _1\sinh 2\iota $

$\widehat{U}^{-1}\left( \kappa \right) \ \Lambda _3\widehat{U}\left( \kappa
\right) =\Lambda _3$

$\widetilde{U}^{-1}\left( \chi \right) \ \Lambda _3\widetilde{U}\left( \chi
\right) =\Lambda _3$

==========

$\widetilde{U}^{-1}\left( \chi \right) \ \Lambda _4\widetilde{U}\left( \chi
\right) =\Lambda _4$

$\widehat{U}^{-1}\left( \kappa \right) \ \Lambda _4\widehat{U}\left( \kappa
\right) =\Lambda _4$

$U_{0,3}^{-1}\left( \iota \right) \ \Lambda _4U_{0,3}\left( \iota \right)
=\Lambda _4\cosh 2\iota -\Lambda _2\sinh 2\iota $

$U_{1,2}^{-1}\left( \varsigma \right) \ \Lambda _4U_{1,2}\left( \varsigma
\right) =\Lambda _4\cos 2\varsigma +\Lambda _1\sin 2\varsigma $

$U_{1,3}^{-1}\left( \vartheta \right) \ \Lambda _4U_{1,3}\left( \vartheta
\right) =\Lambda _4$

$U_{2,3}^{-1}\left( \alpha \right) \ \Lambda _4U_{2,3}\left( \alpha \right)
=\Lambda _4\cos 2\alpha -\Lambda _6\sin 2\alpha $

$U_{0,1}^{-1}\left( \varpi \right) \ \Lambda _4U_{0,1}\left( \varpi
\right) =\Lambda _4\cosh 2\varpi +\Lambda _5\sinh 2\varpi $

==========

$U_{0,1}^{-1}\left( \varpi \right) \ \Lambda _5U_{0,1}\left( \varpi
\right) \ =\Lambda _5\cosh 2\varpi +\Lambda _4\sinh 2\varpi $

$U_{2,3}^{-1}\left( \alpha \right) \ \Lambda _5U_{2,3}\left( \alpha \right)
\ =\Lambda _5\cos 2\alpha +\Lambda _3\sin 2\alpha $

$U_{1,3}^{-1}\left( \vartheta \right) \ \Lambda _5U_{1,3}\left( \vartheta
\right) \ =\left( \Lambda _5\cos 2\vartheta -\Lambda _2\sin 2\vartheta
\right) $

$U_{0,2}^{-1}\left( \phi \right) \ \Lambda _5U_{0,2}\left( \phi \right) \
=\Lambda _5\cosh 2\phi -\Lambda _1\sinh 2\phi $

$U_{0,3}^{-1}\left( \iota \right) \ \Lambda _5U_{0,3}\left( \iota \right) \
=\Lambda _5$

$\widehat{U}^{-1}\left( \kappa \right) \ \Lambda _5\widehat{U}\left( \kappa
\right) \ =\Lambda _5$

$\widetilde{U}^{-1}\left( \chi \right) \Lambda _5\widetilde{U}\left( \chi
\right) \ =\Lambda _5$

===========

$\widetilde{U}^{-1}\left( \chi \right) \Lambda _6\widetilde{U}\left( \chi
\right) \ =\Lambda _6$

$\widehat{U}^{-1}\left( \kappa \right) \Lambda _6\widehat{U}\left( \kappa
\right) \ =\Lambda _6$

$U_{1,2}^{-1}\left( \varsigma \right) \Lambda _6U_{1,2}\left( \varsigma
\right) \ =\Lambda _6$

$U_{0,2}^{-1}\left( \phi \right) \Lambda _6U_{0,2}\left( \phi \right) \
=\Lambda _6\cosh 2\phi +\Lambda _2\sinh 2\phi $

$U_{1,3}^{-1}\left( \vartheta \right) \Lambda _6U_{1,3}\left( \vartheta
\right) \ =\Lambda _6\cos 2\vartheta -\Lambda _1\sin 2\vartheta $

$U_{2,3}^{-1}\left( \alpha \right) \Lambda _6U_{2,3}\left( \alpha \right) \
=\Lambda _6\cos 2\alpha +\Lambda _4\sin 2\alpha $

$U_{0,1}^{-1}\left( \varpi \right) \Lambda _6U_{0,1}\left( \varpi
\right) =\Lambda _6\cosh 2\varpi -\Lambda _3\sinh 2\varpi $

========

$\widetilde{U}^{-1}\left( \chi \right) \ \Lambda _7\widetilde{U}\left( \chi
\right) =\Lambda _7$

$U_{0,3}^{-1}\left( \iota \right) \ \Lambda _7U_{0,3}\left( \iota \right)
=\Lambda _7$

$U_{1,2}^{-1}\left( \varsigma \right) \ \Lambda _7U_{1,2}\left( \varsigma
\right) =\Lambda _7$

$U_{0,2}^{-1}\left( \phi \right) \ \Lambda _7U_{0,2}\left( \phi \right)
=\Lambda _7$

$U_{1,3}^{-1}\left( \vartheta \right) \ \Lambda _7U_{1,3}\left( \vartheta
\right) =\Lambda _7$

$U_{2,3}^{-1}\left( \alpha \right) \ \Lambda _7U_{2,3}\left( \varpi
\right) =\Lambda _7$

$U_{0,1}^{-1}\left( \varpi \right) \ \Lambda _7U_{0,1}\left( \varpi
\right) =\Lambda _7$

=========

$U_{0,1}^{-1}\left( \varpi \right) \ \Lambda _8U_{0,1}\left( \varpi
\right) =\Lambda _8$

$U_{2,3}^{-1}\left( \alpha \right) \ \Lambda _8U_{2,3}\left( \alpha \right)
=\Lambda _8$

$U_{1,3}^{-1}\left( \vartheta \right) \ \Lambda _8U_{1,3}\left( \vartheta
\right) =\Lambda _8$

$U_{0,2}^{-1}\left( \phi \right) \ \Lambda _8U_{0,2}\left( \phi \right)
=\Lambda _8$

$U_{1,2}^{-1}\left( \varsigma \right) \ \Lambda _8U_{1,2}\left( \varsigma
\right) =\Lambda _8$

$U_{0,3}^{-1}\left( \iota \right) \ \Lambda _8U_{0,3}\left( \iota \right)
=\Lambda _8$

$\widehat{U}^{-1}\left( \kappa \right) \ \Lambda _8\widehat{U}\left( \kappa
\right) =\Lambda _8$

\end{document}